%% file: main.tex
\documentclass[fleqn,usenatbib]{mnras}

\usepackage{newtxtext,newtxmath}

\usepackage[T1]{fontenc}

\DeclareRobustCommand{\VAN}[3]{#2}
\let\VANthebibliography\thebibliography
\def\thebibliography{\DeclareRobustCommand{\VAN}[3]{##3}\VANthebibliography}

\usepackage[caption=false]{subfig}
\usepackage[table]{xcolor}
\usepackage{cleveref}

\usepackage{natbib}

\usepackage{graphicx}	
\usepackage{amsmath}	

\input{commands}

\title{Nonlinear Saturation of the Acoustic Resonant Drag Instability}

\author[B.Y. Israeli et. al.]{
Ben Y. Israeli,$^{1,2}$\thanks{E-mail: ben.israeli@weizmann.ac.il}
Jonathan Squire,$^{3}$
Eric Moseley$^{2,4}$
and Amitava Bhattacharjee$^{2}$
\\
$^{1}$Department of Physics of Complex Systems, Weizmann Institute of Science, Rehovot 7610001, Israel\\
$^{2}$Department of Astrophysical Sciences, Princeton University, Princeton, NJ 08540, USA\\
$^{3}$Department of Physics, University of Otago, Dunedin 9016, New Zealand\\
$^{4}$Kavli Institute for Particle Astrophysics \& Cosmology (KIPAC), Stanford University, Stanford, CA 94305, USA
}

\date{Accepted XXX. Received YYY; in original form ZZZ}

\pubyear{2026}

\begin{document}
\label{firstpage}
\pagerange{\pageref{firstpage}--\pageref{lastpage}}
\maketitle

\begin{abstract}
Resonant drag instabilities (RDIs) are a novel type of dust/fluid instability relevant to a diverse range of astrophysical environments.
They are driven by a resonant interaction between streaming dust and waves in a background medium, which results in dust density fluctuations and amplification of the waves.
This broad class of instabilities includes recently-proposed modes incorporating acoustic and magnetohydrodynamic waves, as well as the well-studied disk streaming instability.
As the study of RDIs is at an early stage, their evolution beyond the linear regime is not well understood.
In order to make inroads into the nonlinear theory of RDIs, we performed simulations of the simplest case, the acoustic RDI, in which sound waves in a gas are amplified by interaction with supersonically streaming dust.
This particular instability is of interest both due its potential relevance in various poorly ionized environments, and due to its resemblance to the fast magnetosonic RDI.
We find that the nonlinear growth and saturation of the instability are characterized by a balance between time scales of instability growth and turbulent eddy turnover.
The simulations demonstrate a saturated state possessing an anisotropic outer forcing range in which this balance is maintained, and suggest the presence of an isotropic turbulent inertial range below this scale.
By presenting a model for the nonlinear growth and saturated state of the acoustic RDI, this work provides a framework for further study of the nonlinear behavior of this and other RDIs.
\end{abstract}




\input{sections/1intro}
\input{sections/2simulation}
\input{sections/3saturation}
\input{sections/4stats}
\input{sections/5outer}
\input{sections/6inertial}
\input{sections/7limits}
\input{sections/8conclusion}
\input{sections/9acknowledgments}



\bibliographystyle{mnras}
\bibliography{bibliography} 




\appendix

\input{sections/Abenchmark}
\input{sections/Bspectra}


\bsp	
\label{lastpage}
\end{document}

%% file: commands.tex
\newcommand{\dt}{\partial_t}
\newcommand{\dg}{\cdot\nabla}
\renewcommand{\div}{\nabla\cdot}

\newcommand{\rhod}{{\rho_d}}

\newcommand{\bu}{\mathbf{u}}
\newcommand{\bv}{\mathbf{v}}
\newcommand{\ba}{\mathbf{a}}
\newcommand{\bk}{\mathbf{k}}
\newcommand{\bw}{\mathbf{w}}

\newcommand{\br}{\mathbf{r}}
\newcommand{\bl}{\mathbf{l}}

\newcommand{\sT}{\mathsf{T}}
\newcommand{\sA}{\mathsf{A}}

\newcommand{\sC}{\mathsf{C}}
\newcommand{\sF}{\mathsf{F}}

\newcommand{\sD}{\mathsf{D}}

\newcommand{\sI}{\mathsf{I}}

\newcommand{\sO}{\mathsf{0}}
\newcommand{\sx}{\mathsf{x}}

\newcommand{\grdi}{\gamma_{\rm{RDI}}}
\newcommand{\geddy}{\gamma_{\rm{eddy}}}

\newcommand{\nbd}{\nobreakdashes-}

\newcommand{\im}{{\rm{Im}}}

\newcommand{\kperp}{{k_\perp}}
\newcommand{\kpar}{{k_{||}}}
\newcommand{\koperp}{k^0_\perp}
\newcommand{\kopar}{k^0_{||}}
\newcommand{\tres}{{\theta_{\rm res}}}

\newcommand{\simh}[1]{hr\nbd$w_s$#1}
\newcommand{\siml}[1]{lr\nbd$w_s$#1}
\newcommand{\simu}[1]{lr\nbd$w_s$#1\nbd$\mu$0.1}

%% file: sections/1intro.tex
\section{Introduction}
\label{sec:intro}

Dust is a ubiquitous feature of astrophysical environments, playing a significant role across length and time scales.
Dust grains control radiative energy transport, absorbing, scattering, and thermally emitting radiation.
The resulting radiation pressure on the dust in turn accelerates it. The dust then collisionally transfers momentum to the surrounding gas, driving flows, instabilities, and turbulence.
The dynamics and evolution of the dust grains themselves is complex, with growth driven by condensation and agglomeration, and destruction by collisions and sputtering.
These processes are of substantial interest, both in understanding the observed properties of dust in various environments and in modelling of phenomena such as planet formation.
This provides ample motivation for picking apart the entangled interactions between dust and gas, and for identifying general features of their coupled dynamics.

Resonant drag instabilities (RDIs) are a class of instabilities driven by the interaction of streaming dust with waves in a background fluid.
They are of relevance to a range of astrophysical environments, in which they are theorized to drive turbulence and the formation of complex structures.
RDIs were introduced as a category of instability by \citet{squireResonantDragInstability2018}, resulting in subsequent papers introducing particular cases of the linear instability involving various wave modes \citep{hopkinsUbiquitousInstabilitiesDust2018,squireResonantDragInstabilities2018,hopkinsResonantDragInstability2018}, and discussing nonlinear evolution and impacts on particular astrophysical systems \citep{seligmanNonlinearEvolutionResonant2019,hopkinsDustWindResonant2022,hopkinsSimulatingDiverseInstabilities2020,squireAcousticResonantDrag2022,squirePhysicalModelsStreaming2020,moseleyNonlinearEvolutionInstabilities2019}.
A particular RDI, caused by the interaction of streaming dust with epicyclic oscillations in a rotating disk, was shown to be equivalent to the  well-studied ``streaming instability'' initially introduced by \citet{youdinStreamingInstabilitiesProtoplanetary2005} \citep{squireResonantDragInstabilities2018}.
While this instability has been studied for two decades, the other flavors of RDI, which drive modes such as magnetosonic, acoustic, and Alfv\'en waves, are novel, and their study is at an early stage. In particular, the nonlinear behavior of these instabilities is not well understood.

The nonlinear evolution of RDIs is generically expected to saturate with the clumping of dust and the production of anisotropic turbulent flows.
Studies thus far indicate that the characteristics of the saturated state are strongly dependent on system parameters and on which particular RDIs are active \citep{hopkinsDustWindResonant2022,moseleyNonlinearEvolutionInstabilities2019,seligmanNonlinearEvolutionResonant2019}.
Understanding these phenomena is of importance for refining models of dust's role in diverse astrophysical environments, its formation and destruction, the dynamics of dust-driven winds, and the formation of planets.

With the aim of providing insights into the fundamental features of the turbulent saturated state produced by RDIs, we have performed simulations of the simplest case, the acoustic resonant drag instability produced by uncharged dust streaming supersonically through homogeneous neutral unmagnetized gas.
This instability is expected to be of relevance in poorly ionized environments in which relatively high (supersonic) dust streaming velocities are present.
In particular, it is theorized to play a role in cool stellar winds, environments surrounding active galactic nuclei, and within giant molecular clouds~\citep{hopkinsResonantDragInstability2018,hopkinsUbiquitousInstabilitiesDust2018,israeliResonantInstabilitiesMediated2023a}.
It is also of interest due to its resemblance to the RDI associated with the fast magnetosonic wave, which is anticipated to be of relevance in a broad range of magnetized environments
~\citep{hopkinsResonantDragInstability2018,hopkinsUbiquitousInstabilitiesDust2018}.

\begin{figure}
    \centering
    \includegraphics[width=\linewidth]{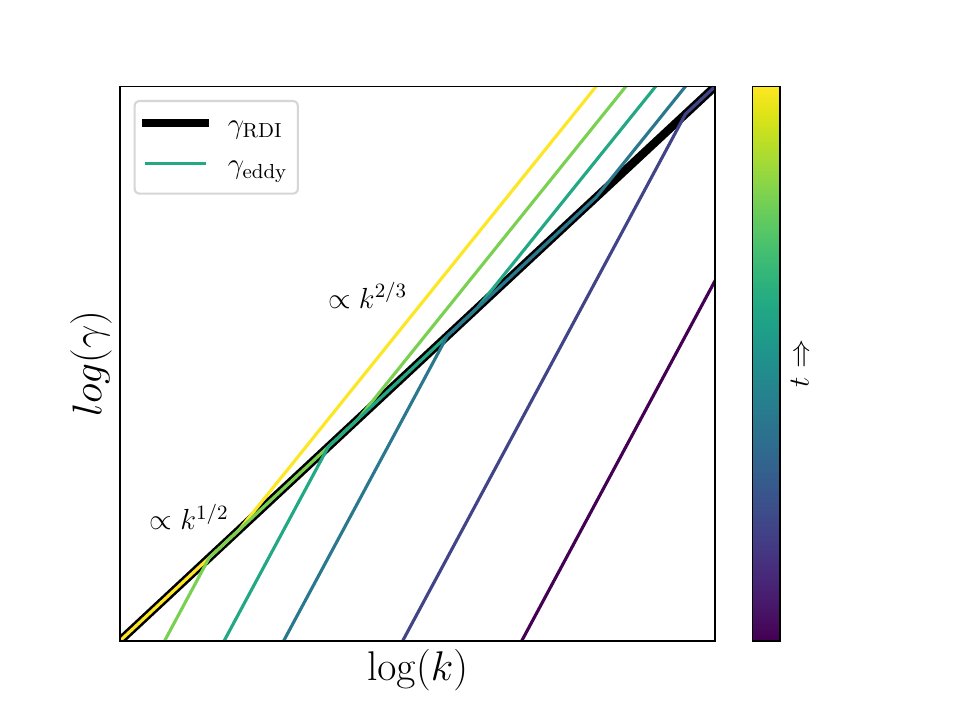}
    \caption{An illustration of the proposed scaling of the eddy turnover rate $\geddy$ as a function of wavenumber $k$ as the system evolves over time $t$ towards a statistically stationary state.
    $\geddy(k)$ for successive times are shown as lines with varied color.
    As discussed in \cref{sec:saturation}, modes initially grow until eddy turnover is in balance with instability growth ($\geddy\sim\grdi\propto k^{1/2}$ shown as a thick black line for reference), reaching this point at smaller scales first.
    In addition, as proposed in \cref{sec:inertial}, at scales sufficiently smaller than the largest saturated scale, a turbulent direct cascade forms ($\geddy\propto k^{2/3}$).
    The resulting inertial range grows until the RDI only remains dominant (and the turnover and growth rates balanced) in a forcing range near the outer scale.
    The resolutions of the simulations in this work resolve the outer range and only the beginning of the inertial range, corresponding to the leftmost portion of this illustration. The range has been extended to smaller scales in order to better demonstrate the scaling in different ranges.}
    \label{fig:schematic}
\end{figure}

Our simulations, performed using the RAMSES code under a variety of conditions, suggest a general model for the nonlinear evolution and saturation of acoustic RDIs.
It is found that the instability saturates first at small scales (at which the instability grows most rapidly), and then at progressively larger scales until reaching the box scale, producing a strongly anisotropic turbulent state.
At each scale, the saturation is due to a balance between linear instability growth and turbulent eddy turnover at that scale.
This balance, originally suggested by \citet{moseleyNonlinearEvolutionInstabilities2019} to be present at the box scale, persists across a range of scales, producing a characteristic gas velocity energy spectrum with a $k^{-2}$ power law.
At smaller scales, we postulate that a turbulent cascade from larger scales may eventually dominate over instability growth, producing progressively more isotropic turbulence.
While the small scales are not sufficiently well-resolved in our simulations, we present evidence consistent with this hypothesis.
Further, the behavior of these simulations is consistent with those performed previously using the GIZMO code, which uses a substantially different architecture, supporting the robustness of both sets of results \citep{moseleyNonlinearEvolutionInstabilities2019}.
In anticipation of the results discussed later in this paper, a schematic description of our proposed model of the evolution and saturation of the acoustic RDI, in terms of the characteristic eddy turnover rate as a function of scale, is given in \cref{fig:schematic}.

In order to give context for our results, the remainder of this section presents a review of acoustic RDIs (\cref{sec:review}) and a summary of prior work on nonlinear acoustic RDIs (\cref{sec:prior}).
Following this, we discuss the code and parameters used in our simulations (\cref{sec:sim}), including comparison to previous work.
Our discussion of the results begins in \cref{sec:saturation}, which concerns the growth and saturation process.
We then move on to the stationary saturated state in \cref{sec:saturated}. \Cref{sec:outer} focuses on the outer scale, at which forcing by the instability is in balance with eddy turnover.
\Cref{sec:inertial} discusses scale-dependent behavior and presents results hinting at the character of the (poorly resolved numerically) inertial range.
In \cref{sec:limits}, the limits in parameter space of the heuristic model used in prior sections are discussed, and mechanisms driving deviation of some simulations from this model are proposed.
A discussion integrating these findings and suggesting further areas of study is provided in \cref{sec:conclusion}.

\subsection{Review of acoustic RDIs}
\label{sec:review}

The streaming of dust through a background medium, often driven by radiation pressure, can act as a source of free energy. (In laboratory experiments, radiation pressure can be applied by external lasers.)
Mediated by the drag the streaming dust imposes on its surroundings, this free energy can drive the amplification of waves within the background medium.
This process, which can drive linear instability when the streaming of the dust tracks the phase velocity $v_p$ of the relevant wave, has been termed the \emph{resonant drag instability}
\citep{squireResonantDragInstabilities2018,squireResonantDragInstability2018,hopkinsUbiquitousInstabilitiesDust2018}.

The linear mechanism driving RDIs, first proposed in \cite{squireResonantDragInstability2018}, may be explained succinctly as follows.
Consider a system consisting of dust streaming through a background fluid with drag between the two.
Schematically, the equations of motion are of the form
\begin{equation}
\label{equ:RDIEoM}
\begin{split}
0=&\dt\rhod+\div(\rhod \bv),\\
0=&\dt \bv+\bv\dg \bv+\frac1t_s(\bv-\bu)-\ba,\\
0=&\dt\rho+\div(\rho \bu),\\
0=&\dt \bu+\bu\dg\bu +\frac{\rho}{\rhod t_s}(\bu-\bv),
\end{split}
\end{equation}
where $\rho$, $\rhod$, $\bu$, $\bv$ are the fluid and dust mass densities and velocities, respectively;
$t_s$ is the drag stopping time of the dust in the fluid,  which may itself be a function of velocities and densities;
$\ba$ is an acceleration applied to the dust but not the gas, radiation pressure for example, which is driving the streaming.
Other terms and equations that capture the dynamics of the fluid, such as pressure, currents, and magnetic fields are ignored because they are not germane in the present context.

Generally the dust constitutes a small fraction of the overall mass density of the system, such that we may take $\mu\equiv\rhod_0/\rho_0\ll 1$.
We may write the linearized system as
\begin{equation}
    i\dot{\sx}=\omega\sx=\sT\sx=(\sT_0+\mu \sT_1)\sx
\label{equ:linear}
\end{equation}
for some state vector $\sx=(\delta \rhod,\delta \bv,\sx_F)$ of the dust density, dust velocity, and fluid state variables (density, velocity, etc.).
Here $\sT_0$ and $\sT_1$  are block matrices:
\begin{equation}
\label{equ:T0T1}
    \sT_0=
    \left(\begin{array}{cc}
            \sA & \sC \\
            \sO & \sF
    \end{array}\right),\ \ 
    \sT_1=
    \left(\begin{array}{cc}
        \sT_1^{AA} & \sT_1^{AF} \\
        \sT_1^{FA} & \sT_1^{FF}
    \end{array}\right).
\end{equation}
The diagonal blocks describe the independent behavior of the dust ($\sA$, $\sT_1^{AA}$) and fluid ($\sF$, $\sT_1^{FF}$) at zeroth and first order in $\mu$, while the off-diagonal blocks describe the influence of the fluid on the dust at zeroth and first order in $\mu$ ($\sC$, $\sT_1^{AF}$), and of the dust on the fluid at first order in $\mu$ ($\sT_1^{FA}$).
For neutral streaming dust in three dimensions,
\begin{equation}
\label{equ:AC}
    \sA=
    \left(\begin{array}{cc}
        \bk\cdot w_s & \bk^T \\
        \sO & \bk\cdot \bw_s \sI+\sD
    \end{array}\right),\ \ 
    \sC=
    \left(\begin{array}{c}
        0  \\
        \sC_\nu 
    \end{array}\right),
\end{equation}
where $\sD$ and $\sC_\nu$ are block matrices describing drag.
The eigenvalue $\omega_0=\bk\cdot \bw_s$ of $\sA$ corresponds to the advection of fluctuations by the streaming dust.
If it resonates with some eigenvalue of $\sF$, corresponding to some oscillatory mode of the fluid, the degeneracy is split by terms in $\sT_1$, yielding at lowest order in perturbation theory
\begin{equation}
\label{equ:RDIgrowth-subbed}
    \omega\approx \bk\cdot \bw_s\pm i\mu^{1/2}\left[(\xi_F^L\sT_1^{(1)})(\bk^T\sD^{-1}\sC_\nu\xi_F^R)\right]^{1/2},
\end{equation}
where $\sT_1^{(1)}$ is the left column of $\sT_1^{FA}$.
(Per \citet{squireResonantDragInstability2018}, for sufficiently large $\bk\cdot \bw_s$, the matrix $\sA$ itself becomes nearly defective, resulting in an overall triply degenerate eigenvalue and an imaginary term which scales as $\mu^{1/3}$.)

Due to the factor of $\pm i$ in equation~\eqref{equ:RDIgrowth-subbed}, a resonance between dust streaming and a wave in the fluid, essentially the dust ``surfing" a wave in the fluid, generically produces a mode with a frequency with positive imaginary part, viz. an instability.
Such a resonance occurs when the streaming velocity exceeds the phase velocity $v_p$ of the wave, appearing at an angle $\theta$ between the wave vector $\bk$ and the dust streaming $\bw_s$ of $\theta=\arccos(v_p/w_s)$. The instability is still active away from this resonance (or with $w_s<v_p$), but attains its maximum growth rate near this \emph{resonant angle}.

Arguably the simplest and most readily understood form of RDI is that of dust coupling to sound waves in a gas.
In this case, the oscillating velocity field of a sound wave clumps dust via drag.
In turn, the resulting fluctuation of the dust density modulates the drag on the gas, amplifying the velocity field of the sound wave.
This process produces a positive feedback loop since, at resonance, the dust and gas fluctuations maintain a constant phase relative to each other \citep{magnanPhysicalPictureAcoustic2024}.

As identified in \citet{hopkinsResonantDragInstability2018} (hereafter HS18), for dust travelling faster than the sound speed in the gas $c_s$, the character of the dominant instability varies with wavelength, fitting into three regimes, listed below. (Here and throughout the rest of the paper, we will reference the (averaged) dust-to-gas mass density ratio $\mu=\left<\rho_d/\rho_g\right>$, and the dust stopping time due to drag (at the initial homogeneous equilibrium) $t_s$.)

\paragraph{Long wavelength/``low-$k$'' $kc_s t_s\ll \mu/(\mu+1)$:}
At long wavelength, the force due to pressure gradients is small compared to drag, such that the system behaves as two drag-coupled pressure-less fluids.
This produces a non-resonant instability whose growth rate scales as $\im(\omega)\propto k^{2/3}$.
It has been termed the ``pressure-free" mode.

\paragraph{Intermediate wavelength/``mid-$k$'' $\mu/(\mu+1) \ll kc_s t_s\ll (\mu+1)/\mu$:}
In this case, the fastest growing mode resembles dust drift ($\omega\approx \bk\cdot \bw_s$), becoming sound-wave-like at the resonant angle.
The growth rate scales as $\im(\omega)\propto k^{1/2}$.
This mode has been termed the ``quasi-drift" mode.

\paragraph{Short wavelength/``high-$k$'' $(\mu+1)/\mu\ll kc_st_s$:}
In this regime, the quasi-drift mode's growth rate instead scales as $\im(\omega)\propto k^{1/3}$.

Away from the resonant angle and at intermediate and shorter wavelengths, an additional mode resembling a sound wave ($\omega\approx kc_s$) becomes relevant.
This mode has a growth rate independent of wavelength and is termed by HS18 the ``quasi-sound" mode.

\begin{figure*}
    \centering
    \includegraphics[width=\linewidth]{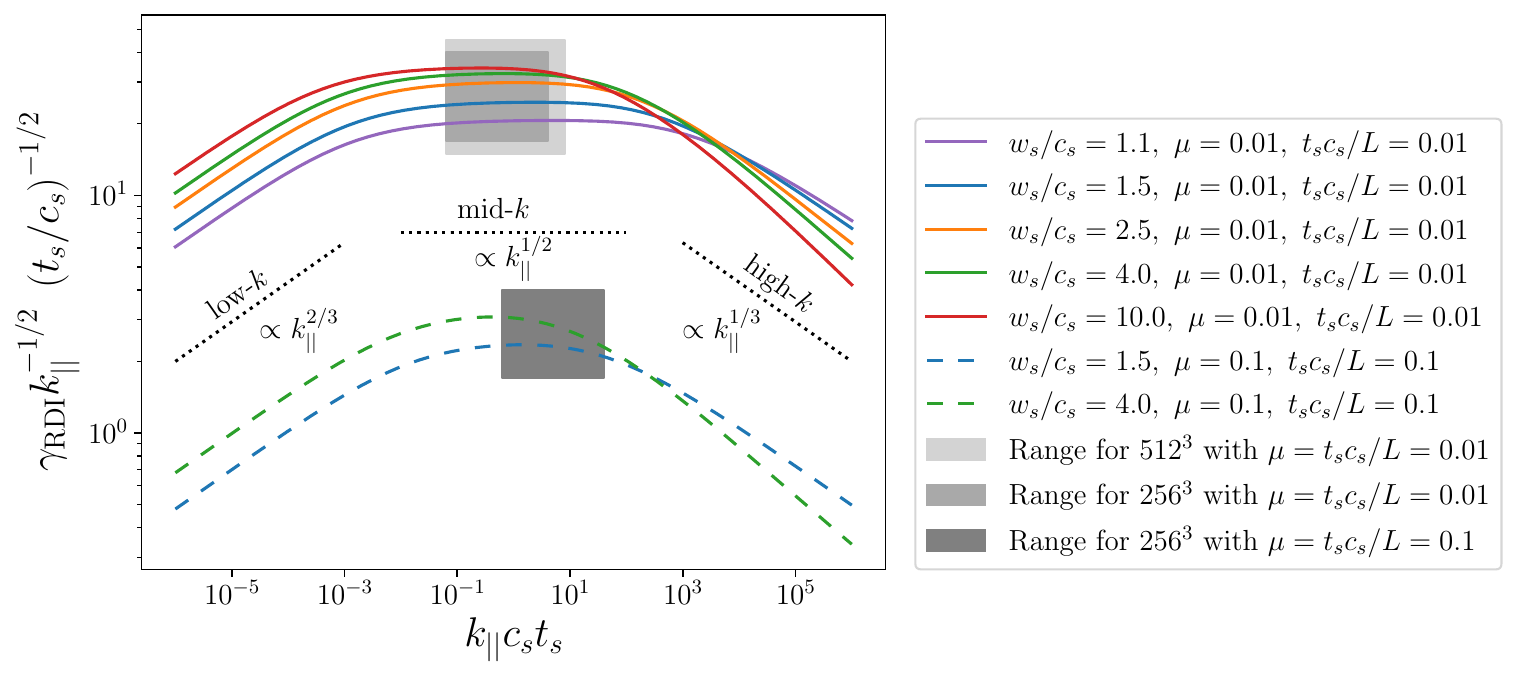}
    \caption{Growth rate of the fastest growing mode at the resonant angle $\grdi$, calculated analytically for the parameters used in each simulation, as a function of wavenumber parallel to streaming $k_{||}$. The growth rate has been multiplied by a factor of $k_{||}^{-1/2}$ to make the power laws for each of the three regimes discussed in \cref{sec:review} clearly visible. Shaded rectangles are shown to indicate the wavelength ranges present in the simulations ($2\pi N/4L\le k_{||}\le 2\pi/L$ for resolution $N$ and box size $L$).}
    \label{fig:analytic}
\end{figure*}

The growth rate of the fastest growing mode at the resonant angle as a function of wavenumber parallel to streaming $k_{||}$ is shown in \cref{fig:analytic} for the parameter values used in the simulations, demonstrating the regimes describe above.

\subsection{Prior work}
\label{sec:prior}
There exists a prior numerical study of the nonlinear saturation of the acoustic RDI, \cite{moseleyNonlinearEvolutionInstabilities2019}.
It presents simulations of monodisperse dust forced through gas in a periodic domain using the GIZMO code \citep{hopkinsGIZMOMultimethodMagnetohydrodynamics2014} with varied stopping time, dust streaming Mach number, and dust mass fraction.
It is found that the instability behavior is determined by its wavelength, divisible into the low-, mid-, and high-$k$ regimes introduced by HS18.
Resonant quasi-drift modes dominate in the mid- and high-$k$ regimes, while the non-resonant pressure-free mode dominates in the low-$k$ regime. By equating the growth rates of HS18 in each of these regimes with the turbulent eddy turnover rate at the box scale, the authors propose heuristic scaling laws for the gas velocity fluctuation amplitude at saturation,
\begin{equation}
\label{equ:moseley}
\frac{\delta u}{c_s}\sim
\begin{cases}
\mu^{1/3}(w_s/c_s)^{2/3}(k_0 c_s t_s)^{-1/3} & k_0c_s t_s\ll \frac{\mu}{\mu+1} \\
\mu^{1/2}(k_0 c_s t_s)^{-1/2} & \frac{\mu}{\mu+1} \ll k_0c_s t_s\ll \frac{\mu+1}{\mu} \\
\mu^{1/3}(k_0 c_s t_s)^{-2/3} & \frac{\mu+1}{\mu}\ll k_0c_st_s,
\end{cases}
\end{equation}
where $k_0$ is the scale of the largest eddies.
They find that their simulations roughly agree with these scalings, and that the linear growth phases of the simulations are also consistent with the the linear growth rates of HS18.
An additional numerical study, focused on the effect of variable dust grain size , has also been performed \citep{squireAcousticResonantDrag2022}.

We verify and extend these findings significantly.
Rather than simply appearing at the largest scales, we find that the balance between eddy turnover and instability growth occurs scale-by-scale across an outer scale range.
We show that the turbulence is strongly anisotropic in this range, and argue, with supporting data, that the turbulence should progressively isotropize below this range.

In addition, it is of note that this study and that of \citet{moseleyNonlinearEvolutionInstabilities2019} obtain consistent results despite substantially different numerical approaches.
\citet{moseleyNonlinearEvolutionInstabilities2019} implement a `meshless finite volume' method for modelling the gas \citep{hopkinsGIZMOMultimethodMagnetohydrodynamics2014} at resolutions of $128^3$ and $256^3$, whereas the fluid equations are solved on a fixed grid at resolutions of $256^3$ and $512^3$ in our simulations.
This suggests that the phenomena observed in these studies are robust and independent of numerical method, and that the behavior of the acoustic RDI has been effectively captured.

%% file: sections/2simulation.tex
\section{Simulations}
\label{sec:sim}

In order to better characterize the physical mechanisms underlying the turbulent saturation of the acoustic RDI, we performed simulations of the instability under varied conditions, focusing on the mid-$k$ regime of the instability.
These simulations were performed using the RAMSES code in a periodic cubic domain.

\subsection{The RAMSES code}
\label{sec:RAMSES}

RAMSES was originally developed as an N-body and hydrodynamics code for cosmological simulations \citep{teyssierCosmologicalHydrodynamicsAdaptive2002}.
It has over two decades of development and use for various astrophysical and cosmological simulations.
The version of RAMSES used in this work implements a magnetohydrodynamic (MHD) fluid model and particle-in-cell (PIC) dust model on a cartesian grid (with dust streaming chosen to be along the $z$-axis)~\citep{moseleyDustDynamicsRamses2023}.
The MHD equations are solved on a discretized cubic domain, with the inclusion of dust back-reaction forces on the fluid.
The dust is modelled via ``superparticles'', which represent collections of identical (same charge, size, and velocity) dust grains distributed within a volume according to a weighting kernel.
The grains evolve according to a momentum equation describing drag, the Lorentz force, and a configurable external force.
This equation is solved via an operator-split semi-implicit method.
For the purposes of this hydrodynamic-PIC study, the magnetic field and the dust charge were set to zero, and the gas was given an isothermal equation of state, such that the equations solved can be written as
\begin{equation}
\label{equ:EoM}
\begin{split}
\frac{\partial \bu}{\partial t}+\bu\cdot\nabla \bu= & -\nabla P+\rho \ba_{\rm{ext},\rm{gas}}
+\int d^3 \bv f_d(\bv)\nu_s(\bv-\bu),\\
\frac{d\bv_i}{dt}= & -\nu_s(\bv_i-\bu)+\ba_{\rm{ext},\rm{dust}},\\
P= & c_s^2\rho.
\end{split}
\end{equation}
Here, $\bu$ and $\bv_i$ are the gas velocity field and dust velocity (for each superparticle separately);
$\rho$, $P$, $\nu_s$, and $f_d(\bv)$ are the gas density, gas pressure, dust drag coefficient, and dust phase space mass density;
$\ba_{\rm{ext},\rm{gas}}$ and $\ba_{\rm{ext},\rm{dust}}$ are the external accelerations applied to the gas and dust, respectively.

This version of the code has been benchmarked and tested for the simulation of RDIs in systems similar to those studied in this work \citep{moseleyDustDynamicsRamses2023}.
An additional set of tests, confirming that the simulation replicates the linear growth rate of an acoustic RDI, and elucidating the resolution limits for realizing such instabilities, is given in \cref{app:benchmark}.
These tests indicate that the simulations accurately capture instabilities with wavelengths longer than approximately $16$ gridpoints, corresponding to $k\approx 100$ or $200$ for $256^3$ and $512^3$ grids, respectively, with instabilities becoming progressively more damped at smaller scales.

\subsection{Simulation parameters}
\label{sec:sim-params}
Simulations were performed in a periodic cubic domain with initially homogeneous and stationary gas and dust.
Dust was forced uniformly through the gas, producing an acoustic RDI.
Each simulation was run until the instability was seen to have saturated for several outer scale eddy turnover times.
All statistics discussed below concerning the saturated state (e.g., spectra, structure functions) have been averaged over the simulation snapshots in this state unless stated otherwise.
    
The box size $L$ and sound speed $c_s$ were set to unity, such that the sound crossing time of the box was unity.
The box resolution was either $256^3$ or $512^3$ cells, and one dust superparticle was initialized at each cell.
Dust drag was treated using an Epstein-Baines model (\citealp[eq. 4]{moseleyDustDynamicsRamses2023}; \citealp{bainesResistanceMotionSmall1965}).
A uniform force along the $z$-axis was applied to the dust, and a counteracting gravitational force was applied to both gas and dust, balanced such that the center of mass velocity of the dust-gas system would remain constant, and the equilibrium streaming velocity $w_s$ of the dust relative to the gas would reach a specified value.
The corresponding formulas for these gravitational and external (applied to the dust) accelerations $g$ and $a$ for a given $w_s$, $t_s$, and $\mu$ are:
\begin{equation}
\begin{split}
    a=&w_s/t_s, \\
    g=&\frac{\mu+1}{\mu}a,
\end{split}
\end{equation}
such that in \cref{equ:EoM}
\begin{equation}
\begin{split}
\ba_{\rm{ext},\rm{gas}}=&-g\hat{z}\\
\ba_{\rm{ext},\rm{dust}}=&(a-g)\hat{z}.
\end{split}
\end{equation}

The RDI growth rate in the targeted mid-$k$ regime scales as $\im(\omega)\sim\mu^{1/2}t_s^{-1/2}$, and the mid-$k$ regime is bounded by $\mu/(1+\mu)\lesssim kc_st_s\lesssim(1+\mu)/\mu$.
As a result, as the dust to gas mass density ratio $\mu$ was varied between runs (taking a value of either $\mu=0.01$ or $\mu=0.1$), the stopping time $t_s$ was set to $t_s=\mu L/c_s$.
This kept the growth rate constant, maintaining the real and simulated runtimes of the simulations, and kept the boundary between mid- and low-$k$ regimes just outside the box scale $k\sim 2\pi/L$.
The boundary between mid- and high-$k$ regimes, meanwhile, moves when $\mu$ and $t_s$ are scaled in this fashion, possibly complicating comparison between runs in which the values of these variables differ.
The changing of $t_s$, with $c_s$ fixed at unity, is equivalent to changing the effective scales captured by the simulation, further complicating comparison.
The wavelength ranges resolvable within the simulation are indicated in \cref{fig:analytic}, showing that the $\mu=0.01$ simulations are firmly in the mid-$k$ regime.
While the $\mu=0.1$ simulations are also largely within this regime, they are near the border between mid-$k$ and high-$k$ regimes.
A list of the simulations of the nonlinear acoustic RDI presented in this paper and their parameters is given in \cref{tab:sims} along with acronyms which will be used for reference.

In addition, simulations of subsonic isotropically driven dust-free turbulence were performed at resolutions of $256^3$ and $512^3$ as a baseline for comparison with the turbulence produced by the acoustic RDI in the other simulations.
These two simulations were performed with identical forcing and other parameters (including an isothermal equation of state), varying only the resolution.
The turbulence was driven by incompressible forcing with a parabolic power spectrum and an auto-correlation time of $0.1\times L/c_s$, resulting in a time-averaged saturated root-mean-squared velocity of $0.27c_s$ and $0.28c_s$ for the $256^3$ and $512^3$ runs, respectively.

\begin{table*}
\centering
\begin{tabular}{|l||c|c|c|c||c|c|c|c|c|}
\hline
name & resolution  & $w_s/c_s$ & $\mu$ & $t_sL/c_s$ 
& $\overline{w_s^{\text{init}}}/c_s$ & $\overline{w_s^{\text{sat}}}/c_s$ & $u_{\text{rms}}/c_s$ & $\rho_{\text{rms}}/\rho_0$ & $\rho_{d\text{rms}}/\rho_{d0}$ \\
\hline
lr\nbd$w_s$1.1 & $256^3$ & 1.1 & 0.01 & 0.01 &
1.08 & 1.13 & 0.33 & 0.066 & 0.039 \\
lr\nbd$w_s$1.5 & $256^3$ & 1.5 & 0.01 & 0.01 &
1.47 & 1.48 & 0.43 & 0.051 & 0.034 \\
lr\nbd$w_s$1.5\nbd$\mu$0.1 & $256^3$ & 1.5 & 0.1 & 0.1 &
1.5 & 1.57 & 0.44 & 0.003 & 0.193 \\
lr\nbd$w_s$2.5 & $256^3$ & 2.5 & 0.01 & 0.01 &
2.43 & 2.48 & 0.54 & 0.028 & 0.033 \\
lr\nbd$w_s$4 & $256^3$ & 4 & 0.01 & 0.01 &
3.86 & 4.01 & 0.73 & 0.025 & 0.046 \\
lr\nbd$w_s$4\nbd$\mu$0.1 & $256^3$ & 4 & 0.1 & 0.1 &
3.99 & 4.14 & 0.66 & 0.015 & 0.406 \\
lr\nbd$w_s$10 & $256^3$ & 10 & 0.01 & 0.01 &
9.85 & 10.08 & 0.78 & 0.045 & 0.079 \\
hr\nbd$w_s$1.5 & $512^3$ & 1.5 & 0.01 & 0.01 &
1.49 & 1.49 & 0.41 & 0.04 & 0.03 \\ 
hr\nbd$w_s$2.5 & $512^3$ & 2.5 & 0.01 & 0.01 &
2.47 & 2.51 & 0.56 & 0.027 & 0.039 \\ 
hr\nbd$w_s$4 & $512^3$ & 4 & 0.01 & 0.01 &
3.94 & 4.04 & 0.71 & 0.025 & 0.064 \\
hr\nbd$w_s$10 & $512^3$ & 10 & 0.01 & 0.01 &
9.93 & 10.1 & 0.85 & 0.025 & 0.14 \\
\hline
\end{tabular}
\caption{Important parameters and statistics of the simulations of the nonlinear acoustic RDI discussed in this paper. Left, varied simulation parameters: resolution, streaming velocity at homogeneous equilibrium $w_s$, overall dust to gas mass density ratio $\mu$, and dust stopping time at homogeneous equilibrium $t_s$. Right, measured values: mass averaged dust streaming velocity $\overline{w_s}=\overline{v}_z-\overline{u}_z$ at the first simulation snapshot with $t>0$ and at saturation, and root-mean-squared fluctuation amplitudes at saturation of the gas velocity $u_{\text{rms}}$, gas density $\rho_{\text{rms}}$, and dust density $\rho_{d\text{rms}}$.}
\label{tab:sims}
\end{table*}

%% file: sections/3saturation.tex
\section{Saturation process}
\label{sec:saturation}
A saturation process common to all simulations was apparent, split into three phases.
Initially, small-scale instabilities, seeded by numerical noise, grew exponentially in the linear regime.
These instabilities quickly saturated, and instabilities and turbulence at progressively larger scales appeared, producing a period of nonlinear growth.
Eventually, a statistically stationary saturated state was produced, with turbulence driven by instabilities at a range of scales.

\subsection{Phases of growth}

\begin{figure}
    \centering
    \subfloat[]{
    \includegraphics[width=\linewidth]{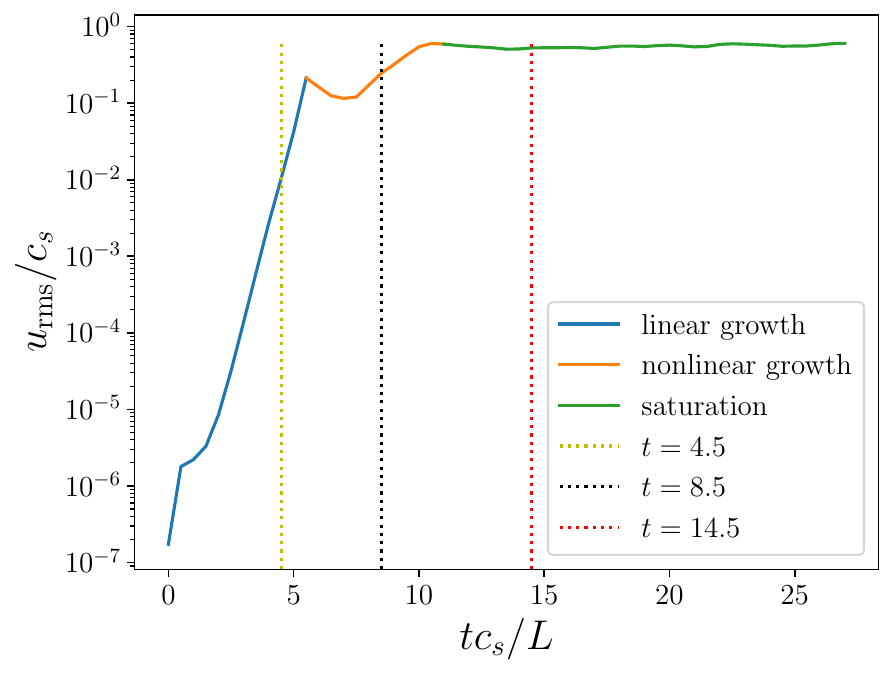}
    \label{fig:u_rms}
    }
    
    \subfloat[]{
    \includegraphics[width=\linewidth]{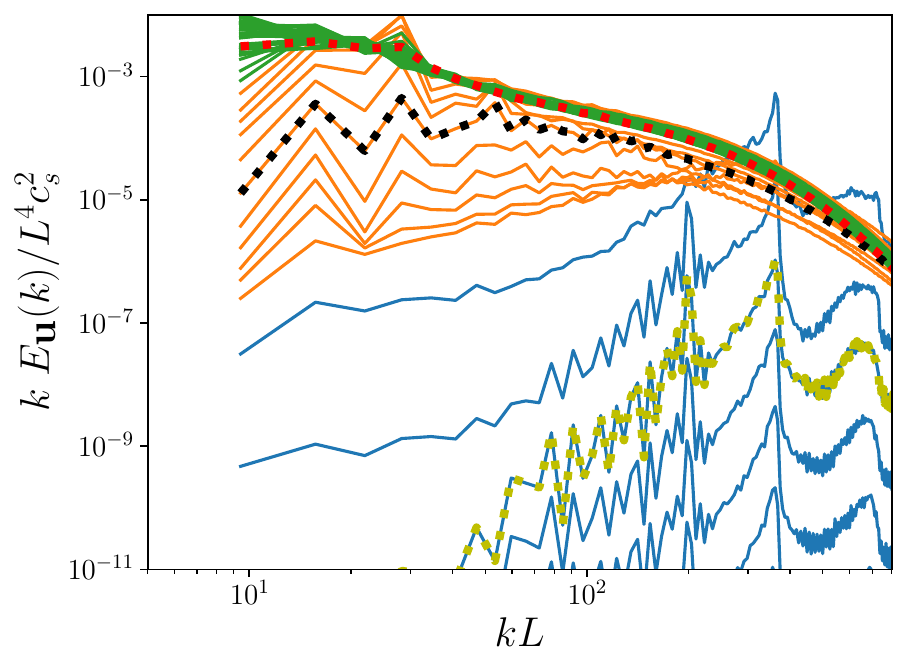}
    \label{fig:u_spec}
    }
    \caption{a) Root mean squared gas velocity for \simh{2.5}, showing the linear growth, nonlinear growth, and saturated phases in blue, orange, and green, respectively.
    b) Gas velocity energy spectra $E_\bu(k)$ for each simulation snapshot, color-coded according to (a).
    Initial growth at small scales, followed by growth and saturation at larger scales, can be seen.}
    \label{fig:u_plots}
\end{figure}

These three phases are visible in the root-mean-squared gas velocity over time in \cref{fig:u_rms}, showing \simh{2.5} for example.
A dip is visible at the start of the nonlinear phase, which was present in all simulations.
It may be attributed to instability ``overshooting'', in the sense that a finite delay is present between the production of turbulent eddies by the instability, and the eddies turning over and halting instability growth.
\Cref{fig:u_spec} shows the evolution of the gas velocity energy spectrum (see \cref{app:spectra}) through these three phases.
Initially, small scale modes are seen to grow, eventually saturating, with a dip in magnitude visible following saturation (as also seen in \cref{fig:u_rms}).
Once the smallest scale modes saturate, the larger scales begin to catch up, saturating from smaller to larger scale.

\begin{figure}
    \centering
    \includegraphics[width=\linewidth]{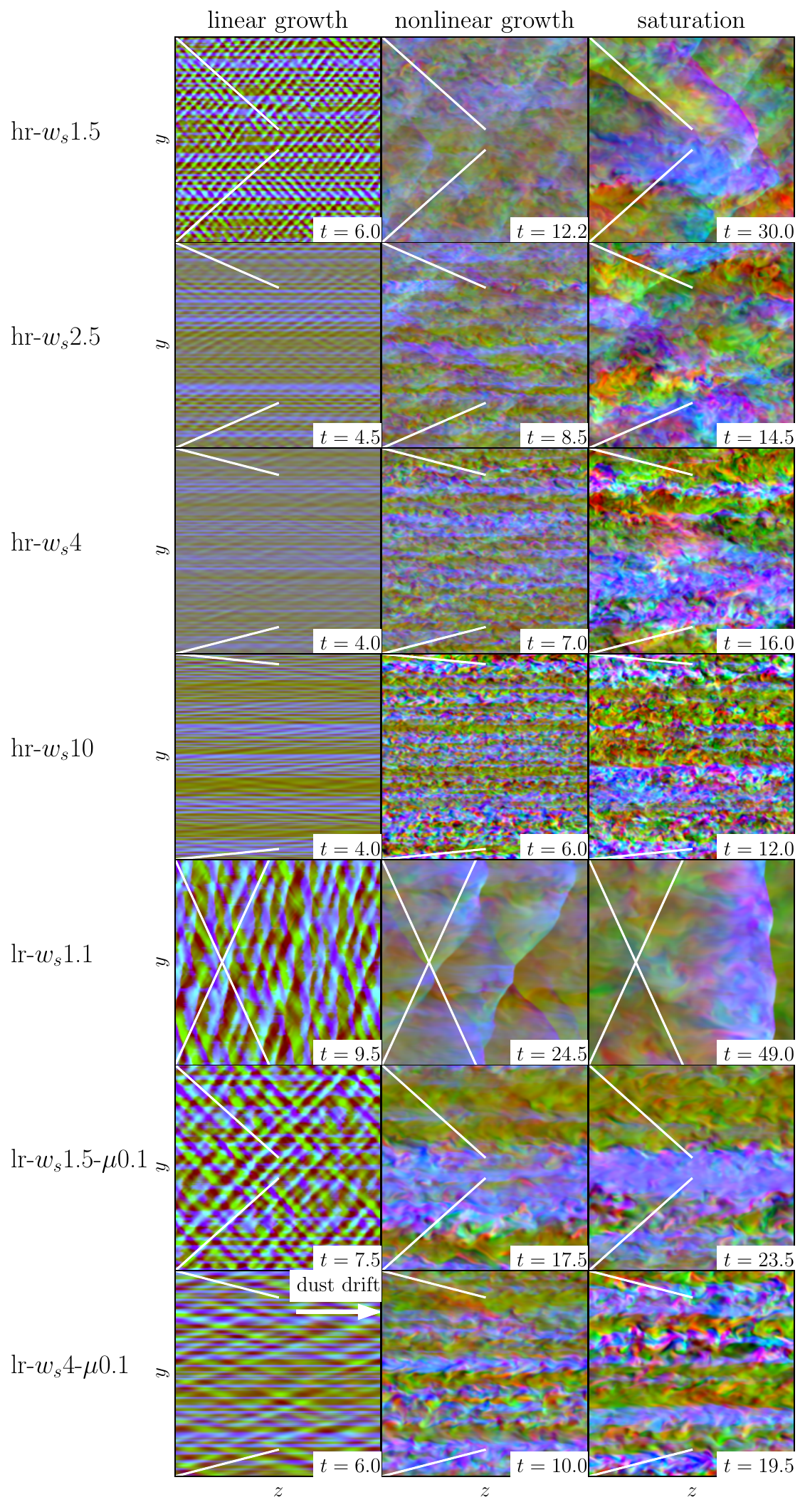}
    \caption{Cross-sections in the $yz$-plane at $x/L=0.5$ of $\bu$ for various simulations, during the linear growth (left), nonlinear growth (center), and saturated phases (right).
    $u_x$, $u_y$, and $u_z$ are depicted using the red, green, and blue channels respectively, normalized to the range $\pm0.025c_s$ (left) or $\pm0.5c_s$ (center, right).
    Ripples at the resonant angle for the instability (indicated with white lines) are visible.
    Times used in each cell are given in units of $L/c_s$.}
    \label{fig:u_cross}
\end{figure}

In \cref{fig:u_cross}, we show images of each phase of evolution in the spatial variation of the gas velocity. One observes small scale instabilities in the linear growth phase, intermediate scale instabilities and small scale turbulence in the nonlinear growth phase, and fully developed turbulence in the saturated state.
It can be seen that ripples that appear to align with the resonant angle of the instability are present even in the saturated state.

\begin{figure}
    \centering
    \includegraphics[width=\linewidth]{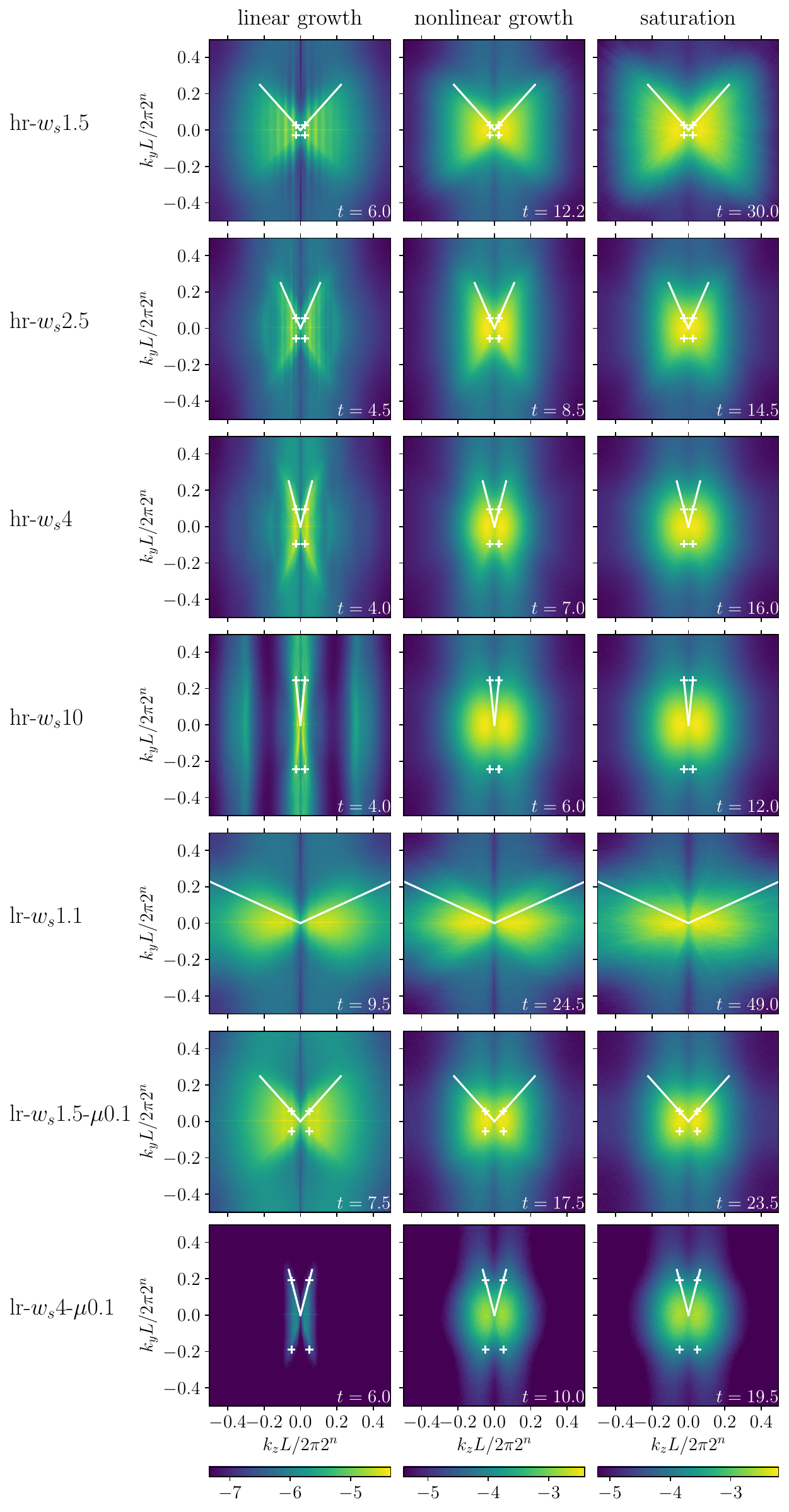}
    \caption{$\log_{10}\left<\tilde\rho^*\tilde\rho/\rho_0^2L^6\right>_{k_y}$ - Gas density fluctuations plotted in Fourier space (averaged over $k_y$) for various simulations, during the linear growth (left), nonlinear growth (center), and saturated (right) phases.
    Cone structures at the resonant angle (indicated with white lines) are visible.
    White crosses are placed at $(k_x,k_z)=(\pm 4\pi\tan(\tres),\pm 4\pi)$ to mark the outer scale as discussed in \cref{sec:outer}.
    Times used in each cell are given in units of $L/c_s$.}
    \label{fig:rho_fourier}
\end{figure}

The presence of resonant modes across a range of scales is further supported by the behavior of gas density fluctuations.
In \cref{fig:rho_fourier}, the magnitude squared of the Fourier transform of the gas density $\tilde\rho(\bk)$ is plotted for the simulations in \cref{fig:u_cross} at each phase of growth, averaged over $k_y$ for ease of plotting.
Cone structures at the resonant angle are visible, indicating the concentration of fluctuations on the surface in momentum space on which resonant instabilities appear.
These are clearly visible during the linear growth phase (left).
During the nonlinear growth phase (center), they persist in most simulations, now concentrated at smaller $k$, and remain present at saturation (right), although in some cases significantly blurred (presumably partially dispersed by turbulence).

\subsection{Scale-by-scale balance}

\begin{figure*}
    \centering
    \subfloat[]{
    \includegraphics[width=0.5\linewidth]{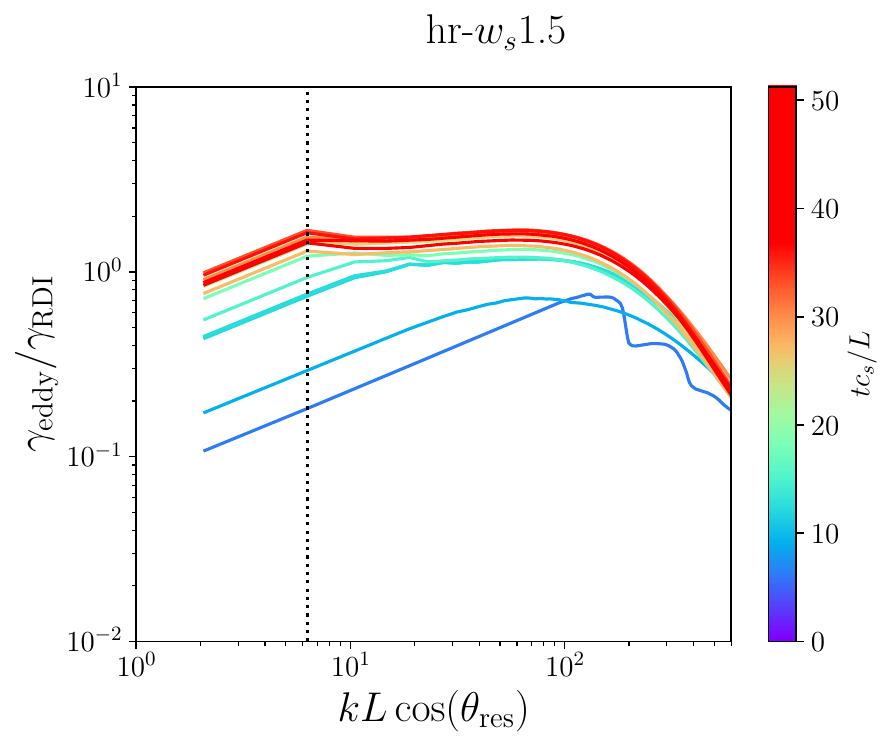}
    }
    \subfloat[]{
    \includegraphics[width=0.5\linewidth]{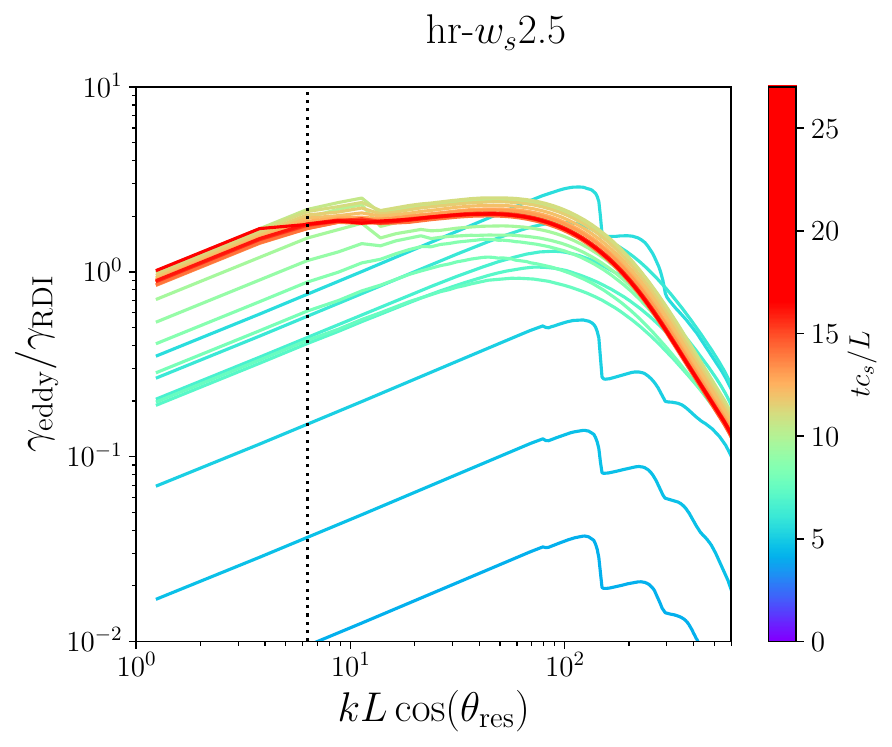}
    }

    \subfloat[]{
    \includegraphics[width=0.5\linewidth]{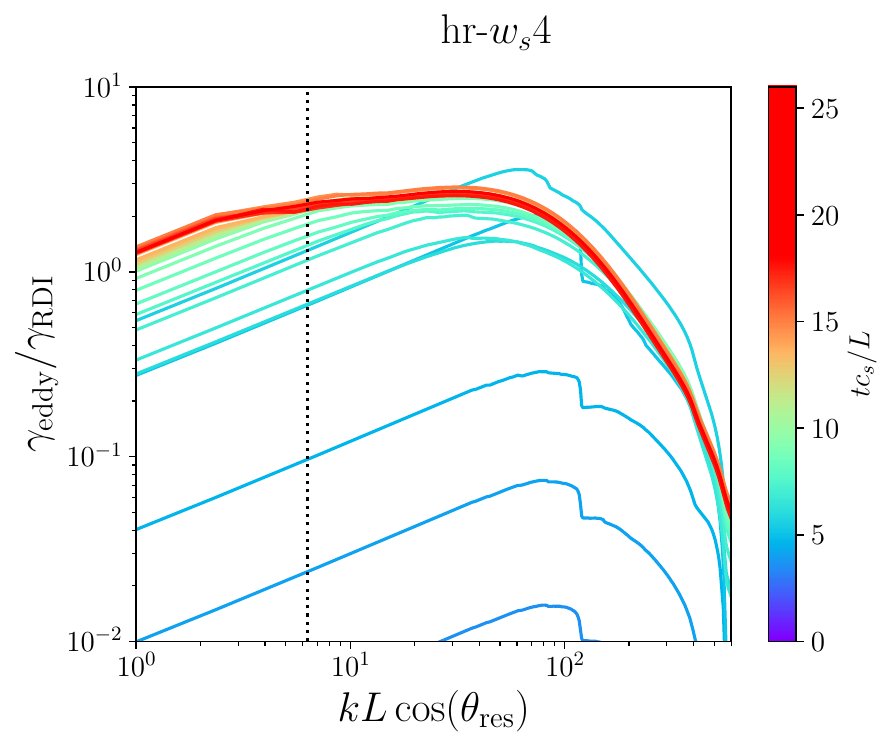}
    }
    \subfloat[]{
    \includegraphics[width=0.5\linewidth]{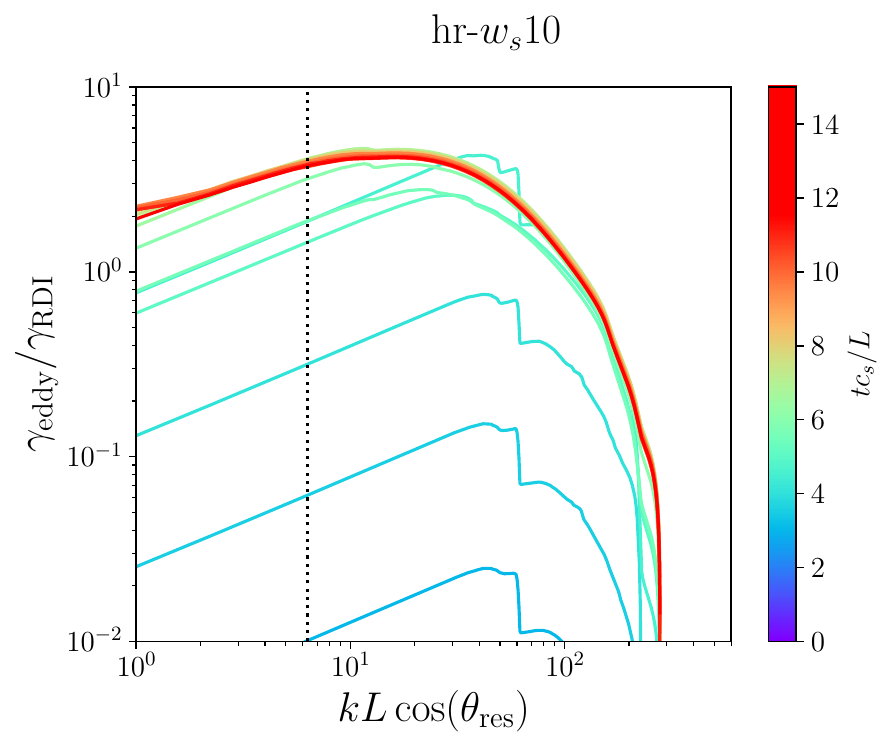}
    }
    \caption{The ratio of the eddy turnover rate (calculated per \cref{app:spectra}) to the linear instability growth rate (calculated analytically) as a function of wavenumber in a series of snapshots from each of the high-resolution simulations.
    A vertical line is shown at $kL\cos(\tres)=2\pi$, the minimum wavenumber that can support a resonant mode in this simulation geometry.
    The observed evolution corresponds to the far left of \cref{fig:schematic}.}
    \label{fig:turnover}
\end{figure*}

By considering the time evolution of the wavelength-dependent ratio of eddy turnover rate to instability growth rate, we find that saturation at each scale  may be understood as a balance between eddy turnover and instability growth.
In a range between the initially growing modes and an outer scale, the instability self-regulates at each scale independently due to the production of turbulent eddies, which disrupt the linearly unstable modes.
The amplitude at which the instability halts is then determined by a balance between the growth and disruption time scales.
With the eddy turnover rate calculated per \cref{app:spectra} and the linear growth rate calculated analytically, the ratio between these rates is plotted in \cref{fig:turnover} for a series of snapshots from each high-resolution simulation.
As already seen in \cref{fig:u_spec}, the smallest scales grow and saturate first, followed by larger scales.
At scales between $kL\cos(\tres)\approx2\pi$, the minimum wavenumber that can support a resonant mode in a cubic periodic domain for $\tres\ge\pi/4$, and $kL\cos(\tres)\approx 10^2$, the ratio saturates to a value of order unity, which varies little with wavenumber (although differing between simulations).
This supports the notion that, in the saturated state, eddy turnover is in balance with instability growth at each scale within an outer forcing range.
This extends the results of \citet{moseleyNonlinearEvolutionInstabilities2019}, who considered this balance only at the box scale.
We provide further evidence in support of this model, in \cref{sec:spectra}, in which we show the gas velocity energy spectra to be consistent with this balance.


\subsection{Nonlinear growth rate}

Assuming this process of saturation by balance between linear and nonlinear processes at progressively larger scales, we may derive an estimate for the time dependence of the nonlinear growth phase. 
Comparison to simulation data yields ambiguous results.

\begin{figure}
\centering
\subfloat[]{
\includegraphics[width=\linewidth]{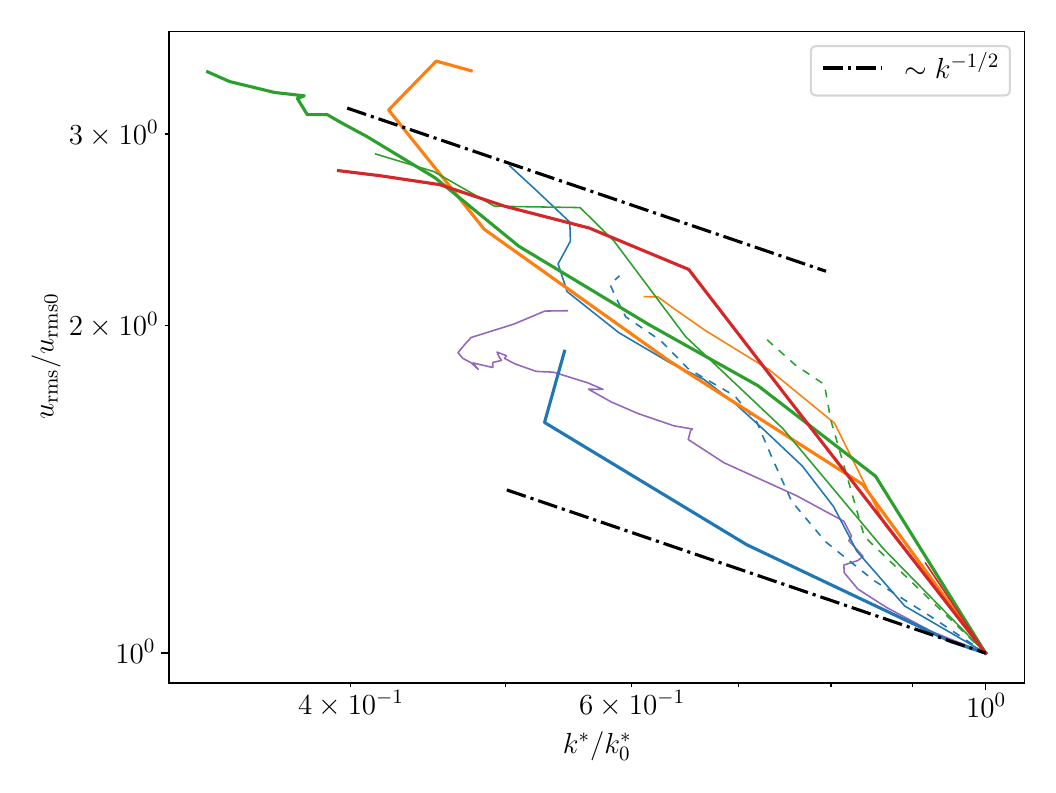}
}

\subfloat[]{
\includegraphics[width=\linewidth]{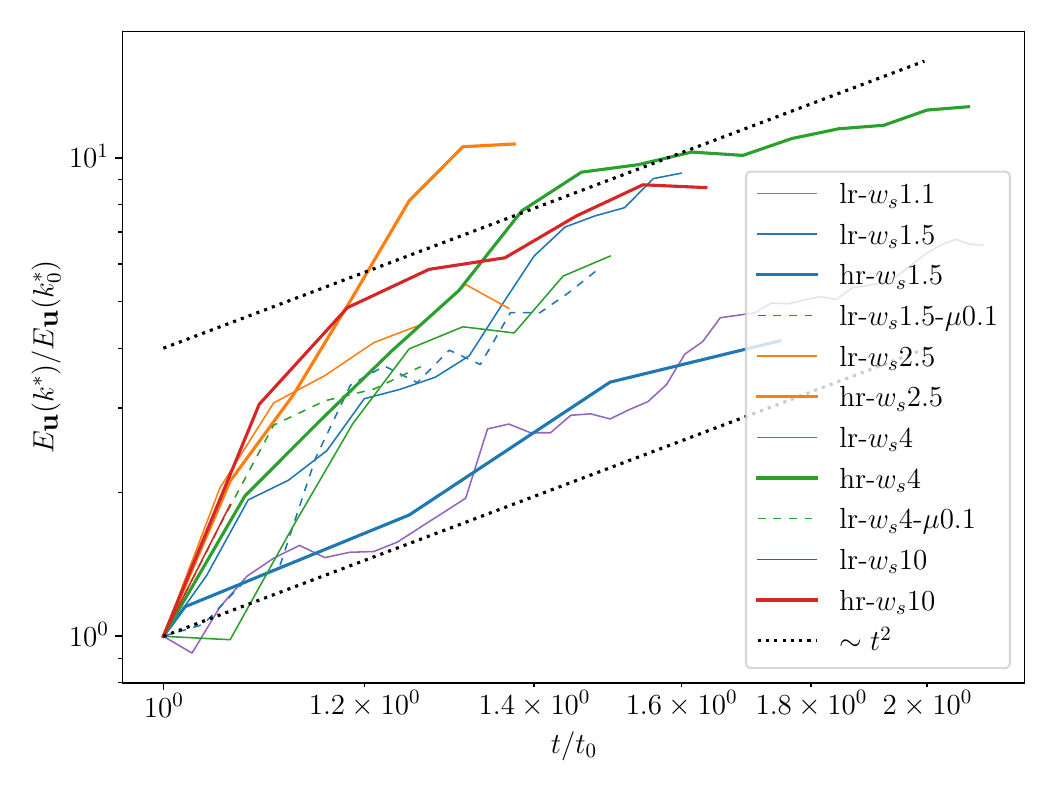}
}

\caption{Testing the scaling relations given by (a) equation~\eqref{equ:uk} and (b) equation~\eqref{equ:et}. $t_0$, $k^*_0$, $u_{\rm rms0}$ denote the values of these variables at the beginning of the nonlinear growth phase.}
\label{fig:turb-nonlin}
\end{figure}

Consider a time $t$ in the nonlinear growth phase where the gas velocity power spectrum is concentrated around a wavelength $k^*(t)$.
As this wavelength saturates at a time around $t$, a balance between the linear growth rate of the instability and the characteristic time of the turbulent flow may be expected at this time.
Using the scaling of the mid-$k$ linear growth rate discussed in \cref{sec:review}, we hypothesize
\begin{equation}
\label{equ:balance}
\grdi\sim {k^*}^{1/2} \sim \geddy \sim u^*k^*,
\end{equation}
where $u^*(t)$ is the characteristic velocity of the flow at the scale $k^*(t)$.
This yields an expectation that $u^*$ should scale as
\begin{equation}
u^*\sim {k^*}^{-1/2}.
\label{equ:uk}
\end{equation}
Driven by the instability (where the turbulence is not yet causing significant dispersion), the energy is expected to grow as
\begin{equation}
\frac{d}{dt}{u^*}^2\sim\gamma {u^*}^2.
\end{equation}
Plugging in equation~\eqref{equ:uk}, this yields
\begin{equation}
\frac{d}{dt}{u^*}^2\sim {k^*}^{1/2}{k^*}^{-1}={k^*}^{-1/2}\sim u^*,
\end{equation}
implying that
\begin{equation}
{u^*}^2\sim t^2.
\label{equ:et}
\end{equation}
These two hypothesized scalings, \cref{equ:uk} and \cref{equ:et}, are tested in \cref{fig:turb-nonlin}.
$k^*$ is measured by a spectrum-weighted average,
\begin{equation}
k^*=\frac{\int dk E_\bu(k) k}{\int dk E_\bu(k)},
\label{equ:kstar}
\end{equation}
where $E_\bu(k)$ is the gas kinetic energy spectrum as defined in equation~\eqref{equ:spec-iso}.
While some simulations seem to agree with one or both of these scaling laws (notably hr\nbd$w_s$10 matching both), there is overall not sufficiently clear agreement to verify this prediction.
The inconsistency of the results may possibly be attributed to the highly approximate nature of $k^*$, which was only heuristically defined in the derivation above and was calculated in a fashion (equation~\eqref{equ:kstar}) that may not correctly capture the largest saturated scale.

%% file: sections/4stats.tex
\section{The saturated state}
\label{sec:saturated}
Having outlined the saturation process and proposed an underlying mechanism, we now discuss the characteristics of the resulting saturated state.

\subsection{Turbulence and its effect on streaming}
\label{sec:stats}
\begin{figure}
    \centering
    \includegraphics[width=\linewidth]{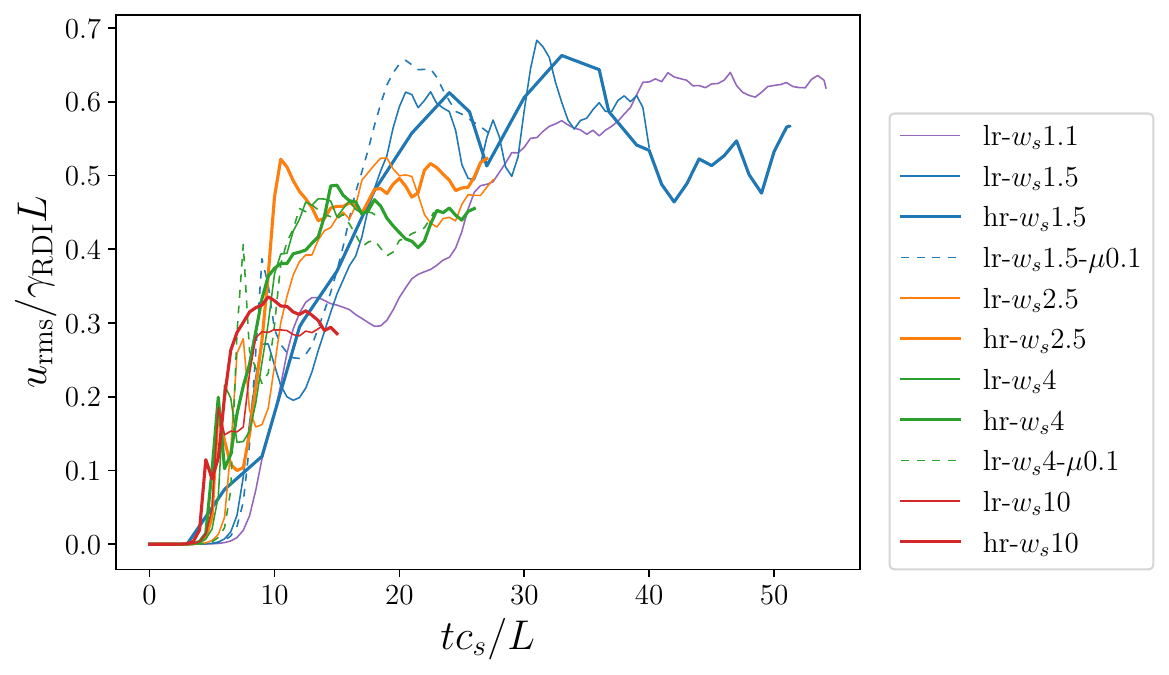}
    \caption{Ratio of root-mean-squared gas velocity to a characteristic velocity given by the box scale and the (analytically calculated) linear growth rate of the acoustic RDI as a function of time for each simulation.}
    \label{fig:urms}
\end{figure}

We begin by commenting on essential features of the saturated state and their dependence on parameters, as listed in the right half of \cref{tab:sims}.
The root-mean-squared fluctuation amplitude of the gas velocity $u_{\text{rms}}$ does not vary significantly with the simultaneous and equal scaling of $\mu$ and $t_s$, consistent with \cref{equ:moseley}, the scaling relation of \cite{moseleyNonlinearEvolutionInstabilities2019}.
As can be seen in \cref{fig:urms}, the scaling postulated by \cite{moseleyNonlinearEvolutionInstabilities2019} that $\delta u\sim u_{\rm rms}\sim \gamma_{\rm RDI} /k_0$ where $k_0$ is the box scale effectively predicts the saturated velocity fluctuation amplitude, with a prefactor of order unity ($\approx0.3-0.6$).
However, the moderate increase of $u_{\rm rms}$ with streaming velocity is not captured by this relation.

In addition, we might expect the presence of turbulence or coherent structures to modify the effective drag coupling between the dust and the gas.
As a proxy for this, we consider the mass-averaged streaming velocity of the dust through the gas, with the force on the dust held constant.
This was measured at the first simulation snapshot taken for which $t>0$ (at which point the instability has yet to grow significantly but after several dust stopping times) $\overline{w_s^{\text{init}}}$, and measured and averaged over the snapshots following saturation $\overline{w_s^{\text{sat}}}$.
While a moderate increase in streaming velocity can be seen in \cref{tab:sims} between the initial and saturated states, the magnitude of this difference is neither consistent with changed resolution nor substantially larger than the variation of these values between the low and high resolution runs.
The results therefore suggest that turbulence or associated clumping of dust may moderately reduce the effective drag coefficient, but are insufficient to clarify the effect in detail.
This runs somewhat counter to intuition applied to Newtonian fluids, as turbulence seems to decrease rather than increase drag.
This may be the result of clumping of dust and subsequent `drafting', as also mentioned by \citet{moseleyNonlinearEvolutionInstabilities2019}.

%% file: sections/5outer.tex
\subsection{Forcing range}
\label{sec:outer}

Returning to the forcing of turbulence by instability, we describe some characteristics of the forcing range at saturation.

\subsubsection{Largest scale modes}
\label{sec:largest}
\begin{figure*}
    \centering
    \includegraphics[width=0.8\linewidth]{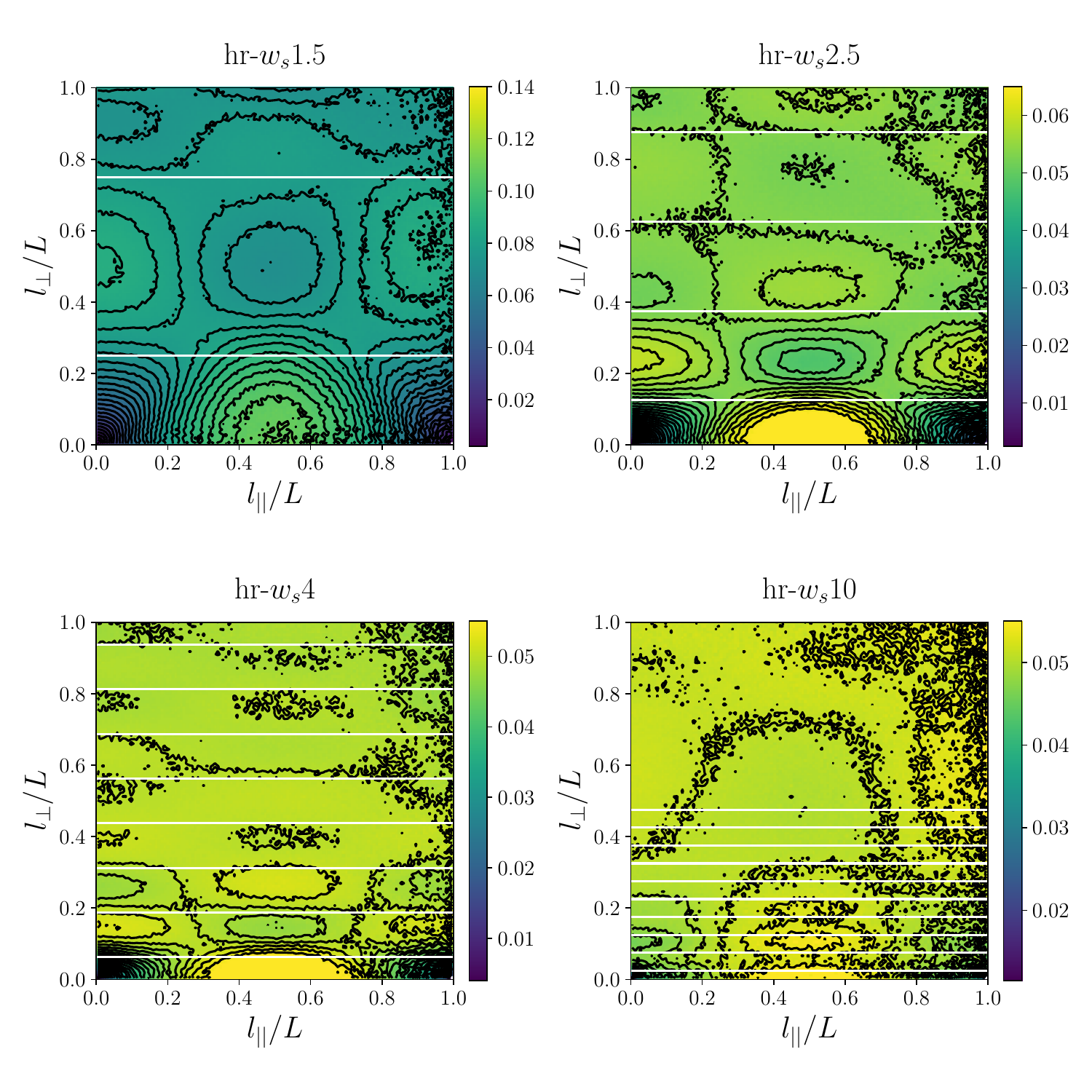}
    \caption{$S_2^{\rho}(l_\perp,l_{||})$, as defined in \cref{equ:s22drho}, for the high resolution simulations.
    This was calculated by random sampling of pairs of points within the simulation domain.
    White lines indicate node lines for the expected outer scale perpendicular wave number as described in \cref{sec:largest}.}
    \label{fig:gas}
\end{figure*}

First, the presence in the saturated state of resonant modes at the largest scales is detectable in the gas density statistics.
For $w_s/c_s\ge\sqrt{2}$, the resonant angle is $\tres\ge\pi/4$, such that the largest scale resonant modes have $\kopar L=2\pi$, which corresponds to $\koperp L=2\pi\tan(\theta_{\rm res})=2\pi\sqrt{w_s^2/c_s^2-1}$.
Structures consistent with these modes are visible in the two-dimensional second-order structure function of the gas density
\begin{equation}
\label{equ:s22drho}
S_2^{\rho}(l_\perp,l_{||})=\left<\left[\rho\left(\br+l_{\perp}\hat{\bl}_\perp+l_{||}\hat{\mathbf{z}}\right)-\rho(\br)\right]^2\right>_\phi,
\end{equation}
where an average has been taken over all $\br$ and over the azimuthal angle $\phi$ of the unit vector $\hat{\textbf{l}}_\perp$, which has $\hat{\textbf{l}}_\perp\cdot\hat z=0$.
This is plotted in \cref{fig:gas} for the $512^3$ simulations.
An array of nodes and antinodes is visible.
We see that indeed $\kopar L\approx 2\pi$.
For $w_s/c_s=1.5,\ 2.5,\ 4,\ 10$, we have $\tan(\theta_{\rm res})=1.12,\ 2.29,\ 3.87,\ 9.94$, suggesting that $\koperp L/2\pi=1,\ 2,\ 4,\ 10$ is the closest available mode matching the periodic boundary conditions.
The node lines for this $\koperp$ are shown in white.
\simh{1.5}, \simh{2.5}, and \simh{4} are consistent with this prediction, while the $\koperp$ for \simh{10} seems to be somewhat lower.
This may be explained by the increased robustness of larger wavelength modes outweighing the higher growth rate of modes nearer resonance, or perhaps by the retardation of smaller wavelength modes by numerical dissipation, as discussed in Appendix~\ref{app:benchmark}.

\subsubsection{Energy spectra}
\label{sec:spectra}
\begin{figure*}
    \centering
    \includegraphics[width=0.9\linewidth]{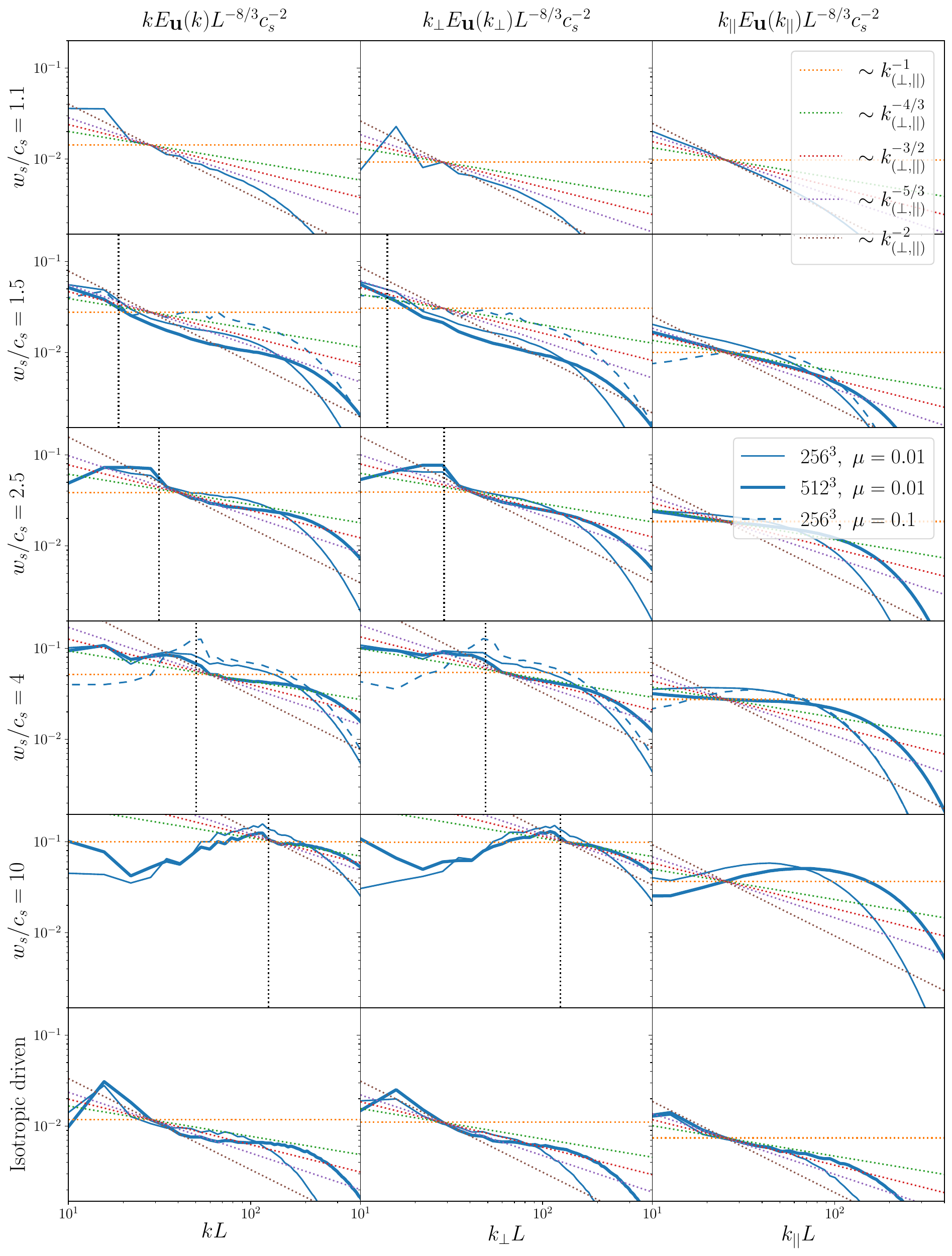}
\caption{Compensated gas velocity power spectra as defined in \cref{equ:spec-iso,equ:spec-aniso} for the all simulations, taken with respect to $k$, $k_{\perp}$, and $k_{||}$.
The spectra are multiplied by $k_{(\perp,||)}$ in order to make comparison of the slopes to the shown power laws clearer.
The $\mu=tc_s/L=0.01$ simulations at resolutions of $256^3$ and $512^3$ are shown with thin and thick lines respectively; the $\mu=tc_s/L=0.1$ simulations at a resolution of $256^3$ are shown with dashed lines.
For visual comparison, dotted reference lines with different power laws are shown in various colors.
A vertical dashed line is used to demarcate $k^0=4\pi w_s/c_s$ or $k^0_\perp=4\pi\sqrt{w_s^2/c_s^2-1}$ as discussed in \cref{sec:spectra}.}
    \label{fig:1d-spectra}
\end{figure*}

The energy spectra of the gas velocity, taken with respect to $k$, $\kpar$, and $\kperp$ (relative to dust streaming direction and calculated per \cref{app:spectra}) are found to describe a strongly anisotropic turbulent state with, in some cases, a visible power law with respect to $\kperp$ determined by the balance between linear and nonlinear times described in \cref{sec:saturation}.
These spectra are plotted in \cref{fig:1d-spectra}.

A knee, defining the upper end of the forcing range, is visible in the spectra with respect to $k$ and $\kperp$ near the expected largest scale resonant modes for many of the simulations.
These are marked with vertical lines indicating the resonant $\koperp$ and $k^0=\sqrt{(\koperp)^2+(\kopar)^2}$ for $\kopar=4\pi$.

The balance described in \cref{sec:saturation} constrains the expected slope of the spectra in the forcing range.
Within the mid-$k$ regime studied in these simulations, the instability growth rate scales as
\begin{equation}
\label{equ:rdi_scaling}
    \grdi\sim k^{1/2}.
\end{equation}
Meanwhile, per \cref{equ:eddy_expression}
\begin{equation}
\label{equ:eddy_scaling}
\geddy(k)\sim k^{3/2} \left[E_\bu(k)\right]^{1/2}.
\end{equation}
Supposing that (for some outer forcing range)
\begin{equation}
    \grdi\sim\geddy,
\end{equation}
and applying \cref{equ:rdi_scaling,equ:eddy_scaling}, we obtain
\begin{equation}
\label{equ:scaling}
E_\bu(k)\sim k^{-2}.
\end{equation}

Reference lines showing the slope $\sim k^{-2}$ have been included in \cref{fig:1d-spectra}.
It can be seen in \simh{1.5} and \siml{1.5} that there is a substantial range near the outer scale where this scaling law is closely followed in the spectra with respect to $k$ and $\kperp$.
A smaller such range may be seen for \simh{2.5} and \siml{2.5}.
The forcing scale width then becomes too small for $w_s/c_s=4,\ 10$ for the slope to be discerned.

The turbulence is driven in a strongly anisotropic fashion, with both the dust streaming direction and the resonant angle of the instability breaking spherical symmetry.
As such, it is not surprising that the parallel and perpendicular spectra have dramatically different behavior.
At large scales, the parallel spectra seem to have constant slopes, which are dependent on streaming velocity.

%% file: sections/6inertial.tex
\subsection{Inertial range and (An)isotropy}
\label{sec:inertial}

The instability-driven turbulence at the forcing scale will  produce a turbulent cascade to smaller scales.
If the eddy turnover rate at smaller scales resulting from this cascade scales with $k$ more strongly than the instability growth rate ($\grdi\sim k^{1/2}$), then the balance present at large scales will not be present at smaller scales.
Instead, the turbulence at smaller scales will be primarily driven by the cascade, with the acoustic RDI suppressed by rapid eddy turnover, producing an inertial range.
The proposed behavior of the characteristic eddy turnover rate as a function of scale as the system evolves towards saturation is illustrated in \cref{fig:schematic}.
Sufficiently far from the forcing scale, the turbulence would be expected to approach universal isotropic behavior, perhaps possessing a Kolmogorov power law, and yielding $\geddy\sim k^{2/3}>k^{1/2}\sim\grdi$, consistent with this argument.
It is worth noting that the presence of this inertial range is dependent on the slow scaling of the growth rate $\grdi\sim k^{1/2}$, and that instabilities with growth rates that scale faster in $k$ than the direct cascade turnover rate would not be likely to produce such behavior, instead remaining in balance across all scales.

As can be seen in \cref{fig:1d-spectra}, while the $k^{-2}$ law associated with forcing and balance seems to break down at smaller scales, the resolution of these simulations is not sufficient to resolve a clear inertial range below the forcing (balanced) scale.
This limitation is also apparent in the test cases with isotropically driven dust-free turbulence.
While a small region in the spectrum with respect to $k$ is present in which the slope of the spectrum matches the expected Kolmogorov $k^{-5/3}$ law, the spectrum is dominated by a forcing range with steeper slope and a region near the dissipation scale with a shallower slope, the latter of which may be attributed to a ``bottleneck effect'' \citep{doblerBottleneckEffectThreedimensional2003,donzisBottleneckEffectKolmogorov2010,schmidtNumericalDissipationBottleneck2006}.
As such, our ability to study the inertial range is limited.

Nonetheless, we discuss the behavior seen over the range available in our simulations and its relevance to this hypothesis.
In particular, it is possible to observe a tendency towards isotropy at smaller scales within the range accessible in these simulations, consistent with an eventual transition to an isotropic inertial range.

\subsubsection{Eddy aspect ratio}
\label{sec:eddies}
\begin{figure*}
\centering
\includegraphics[width=0.8\linewidth]{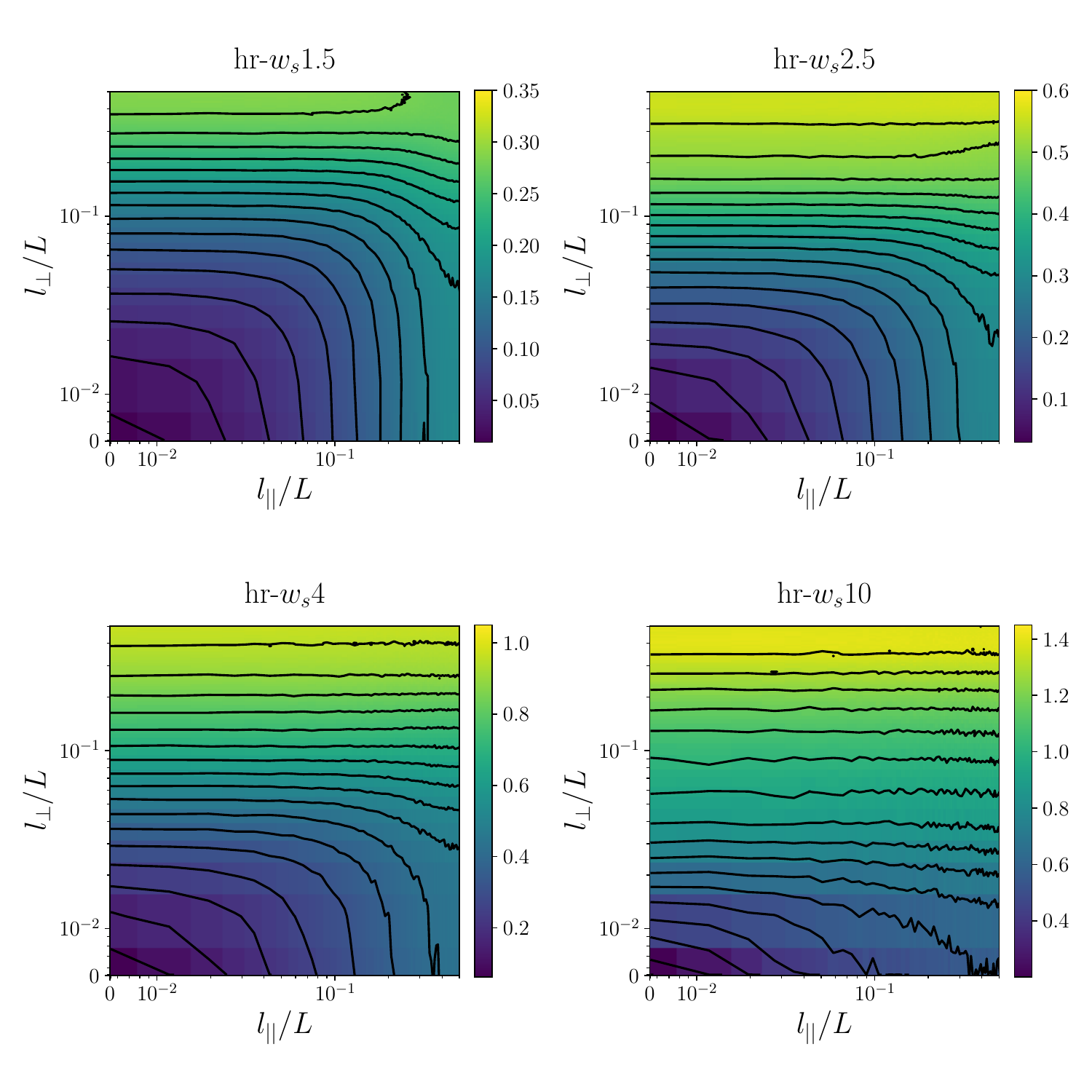}
\caption{$S_2^{\rm{sol}}(l_\perp,l_{||})/c_s^2$ (per \cref{equ:s22d}) for the $512^3$ RDI simulations.
Isocontours act as a proxy for typical eddy shape at different scales.}
\label{fig:eddies}
\end{figure*}

\begin{figure*}
\centering
\includegraphics[width=0.8\linewidth]{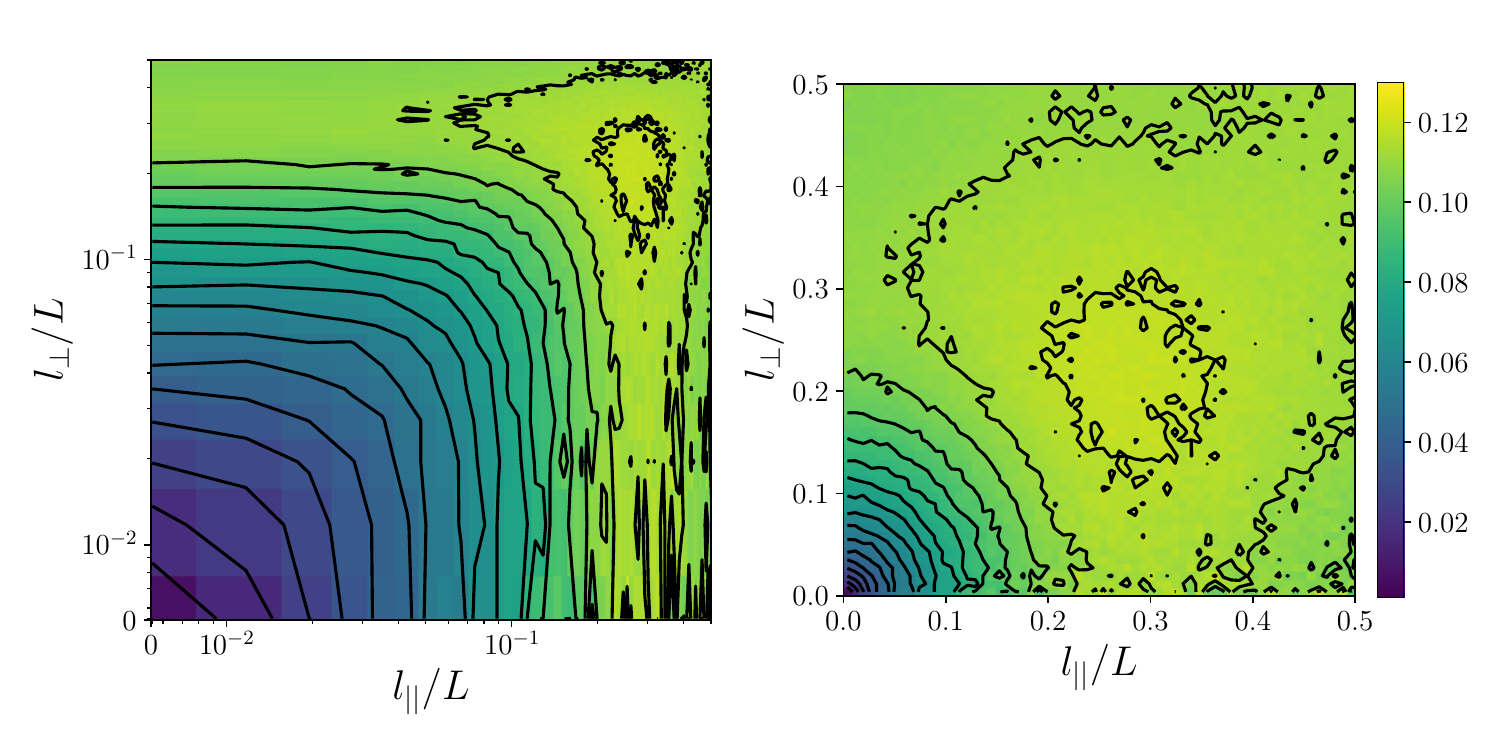}
\caption{$S_2^{\rm{sol}}(l_\perp,l_{||})/c_s^2$ (per \cref{equ:s22d}) for the $512^3$ isotropically driven turbulence simulation.
Isocontours act as a proxy for typical eddy shape at different scales.
It is plotted on a symmetric logarithmic scale  \citep{caswellMatplotlibMatplotlibREL2023} (left) and a linear scale (right) in order to show both the small scale structure and the circularity of the contours.}
\label{fig:kolmogorov}
\end{figure*}

As a proxy for eddy shape, we study isocontours of the two-dimensional second-order structure function of the solenoidal component of the gas velocity, defined as
\begin{equation}
\label{equ:s22d}
S_2^{\rm{sol}}(l_\perp,l_{||})=\left<\left[\bu_{\rm{sol}}\left(\br+l_{\perp}\hat{\bl}_\perp+l_{||}\hat{\mathbf{z}}\right)-\bu_{\rm{sol}}(\br)\right]^2\right>_\phi,
\end{equation}
where (as in \cref{equ:s22drho}) an average has been taken over all $\br$ and over the azimuthal angle $\phi$ of the unit vector $\hat{\textbf{l}}_\perp$, which has $\hat{\textbf{l}}_\perp\cdot\hat z=0$.
This is plotted for the $512^3$ RDI simulations in \cref{fig:eddies}, and, for comparison, for the $512^3$ isotropically driven turbulence simulation in \cref{fig:kolmogorov}.
If an eddy with width $l_\perp$ typically has length $l_{||}$, then the (solenoidal) velocity increment for displacements of those distances and directions should be comparable.

As expected, the isotropically driven turbulence has isotropic, circular eddies.
The acoustic RDI simulations have elongated eddies, whose aspect ratios increase with increasing diameter, becoming shear flows across the simulation box at sufficient separation.
This effect scales dramatically with streaming velocity, with hr\nbd$w_s$2.5 and hr\nbd$w_s$4 having markedly elongated eddies, and hr\nbd$w_s$10 only having eddies at the smallest scales.
Consistent with the eventual dominance of an isotropic cascade, the eddies are progressively more isotropic at smaller scales.

\subsubsection{Two-dimensional spectra}
\label{sec:2d-spectra}
\begin{figure}
\hspace{-0.15\linewidth}\includegraphics[width=1.3\linewidth]{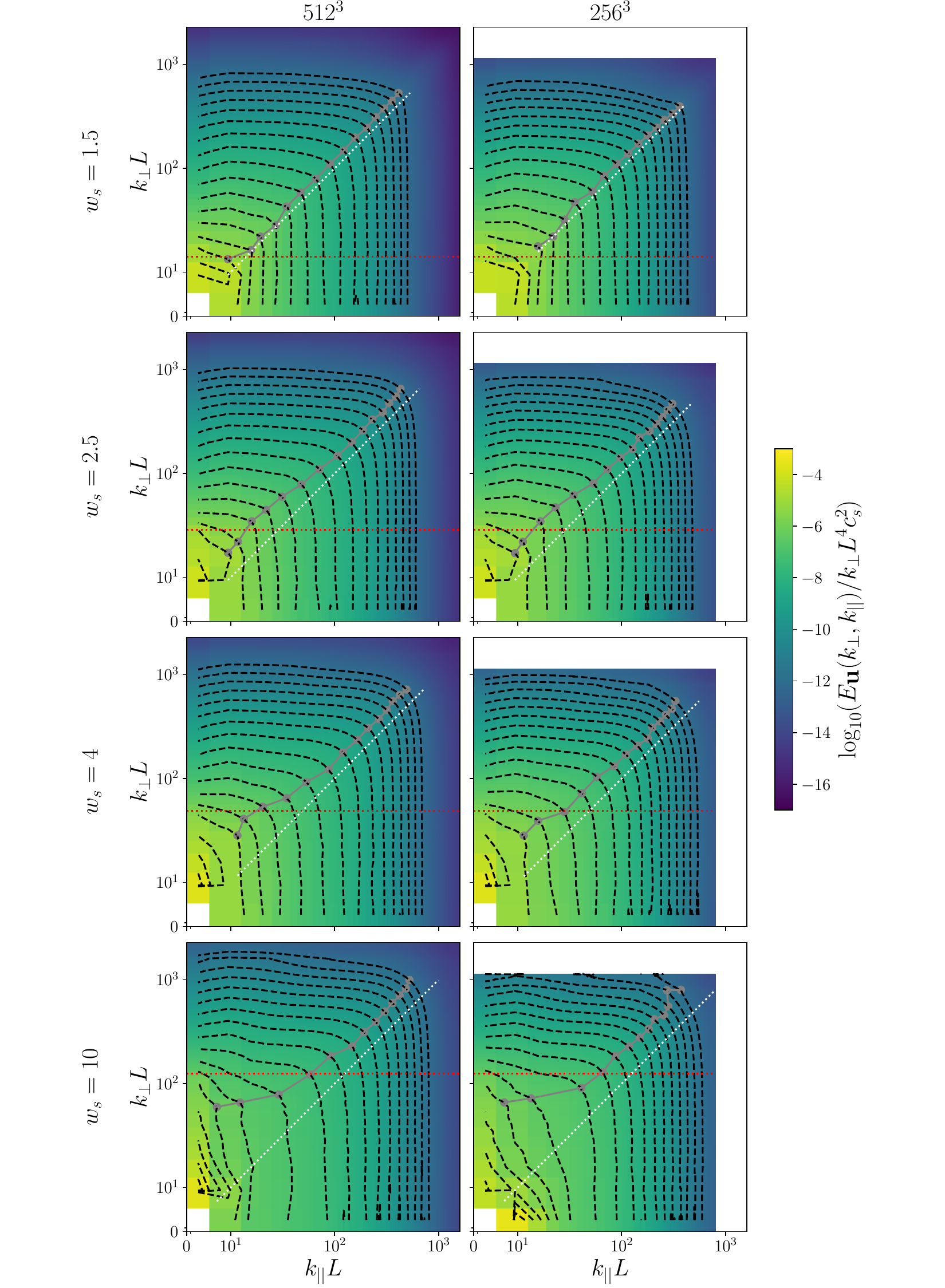}
\caption{Two-dimensional power spectra for several simulations with specified $w_s$ and resolution and with $\mu=0.01$, $tc_s/L=0.01$, divided by $k_\perp$ such that an isotropic spectrum would appear symmetric as described in \cref{sec:2d-spectra}.
Isocontours are shown in black, and the ``ridge" in the distribution is traced in gray by maximizing $k_\perp^2k_{||}^2$ along each contour.
$k_\perp=k_{||}$ is shown in white, and $k^0_\perp L=4\pi\sqrt{w_s/c_s^2-1}$ is shown in red (indicating the forcing range as in \cref{fig:1d-spectra}).}
\label{fig:2d-spectra}
\end{figure}

This transition from anisotropy to isotropy with scale is also visible in the two-dimensional gas velocity energy spectra, defined as
\begin{equation}
E_\bu(k_\perp,k_{||})=\int d\theta k_\perp E_\bu(\bk),\\
\label{equ:spec-2d}
\end{equation}
where $E_\bu(\bk)=\frac12 \left|\tilde \bu(\bk)\right|^2$ is the three-dimensional spectral energy density.
Note that if $E_\bu(\bk)$ is isotropic ($E_\bu(\bk)=\frac1{4\pi}E_\bu(k)$) in three dimensions, then $E_\bu(k_\perp,k_{||})/k_\perp=\frac12 E_\bu(k)$ is isotropic in two dimensions. We plot in \cref{fig:2d-spectra} the ``isotropized'' two-dimensional spectra $E_\bu(k_\perp,k_{||})/k_\perp$ for the $512^3$ runs and their $256^3$ counterparts.

\begin{figure}
\centering
\includegraphics[width=\linewidth]{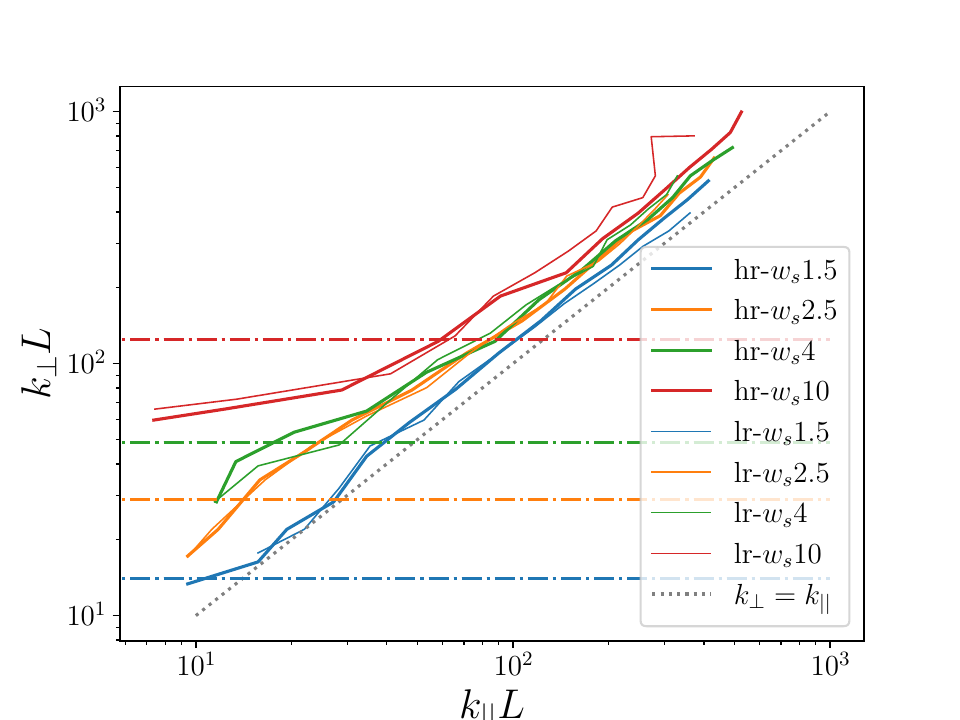}
\caption{The ``ridges" traced in \cref{fig:2d-spectra}.
The line $\kperp=\kpar$ is shown for reference.
The colored horizontal lines show $k^0_\perp L=4\pi\sqrt{w_s/c_s^2-1}$ as also indicated in \cref{fig:1d-spectra,fig:2d-spectra}.
It can be seen that the location of the ridge at large scale is increasingly anisotropic with increasing dust streaming velocity, but trends toward $k_{\perp}\sim k_{||}$ at small scale, consistent with turbulence beginning to dominate over instability.}
\label{fig:ridge}
\end{figure}

The spectra are anisotropic near the outer scale, with anisotropy increasing with dust streaming velocity, as may be expected.
Each spectrum has a distinct ``ridge" along which its isocontours turn sharply.
The location of the turning point in each isocontour was approximated by maximizing $k_\perp^2k_{||}^2$ along each isocontour.
The resulting trace of the ridge is plotted in \cref{fig:ridge}.
It can be seen that while the large scale behavior varies, becoming more anisotropic near the forcing scale with higher streaming velocity, these lines approach a universal isotropic behavior ($k_\perp\sim k_{||}$) at smaller scales.

%% file: sections/7limits.tex
\section{Deviation from the balanced instability model}
\label{sec:limits}
\begin{figure}
    \centering
    \subfloat[]{
    \includegraphics[width=\linewidth]{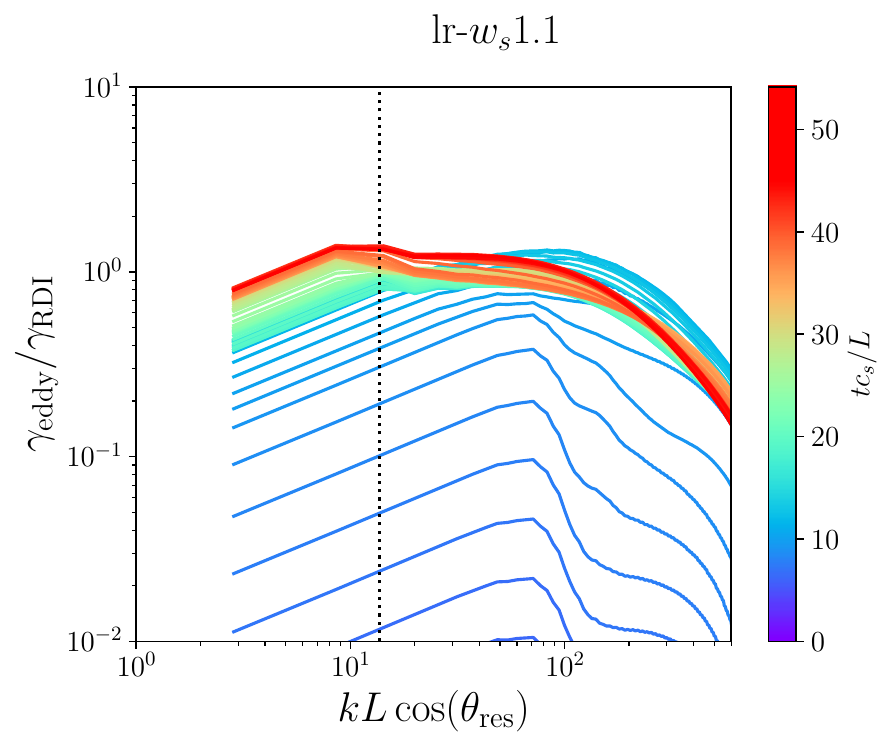}
    }
    
    \subfloat[]{
    \includegraphics[width=\linewidth]{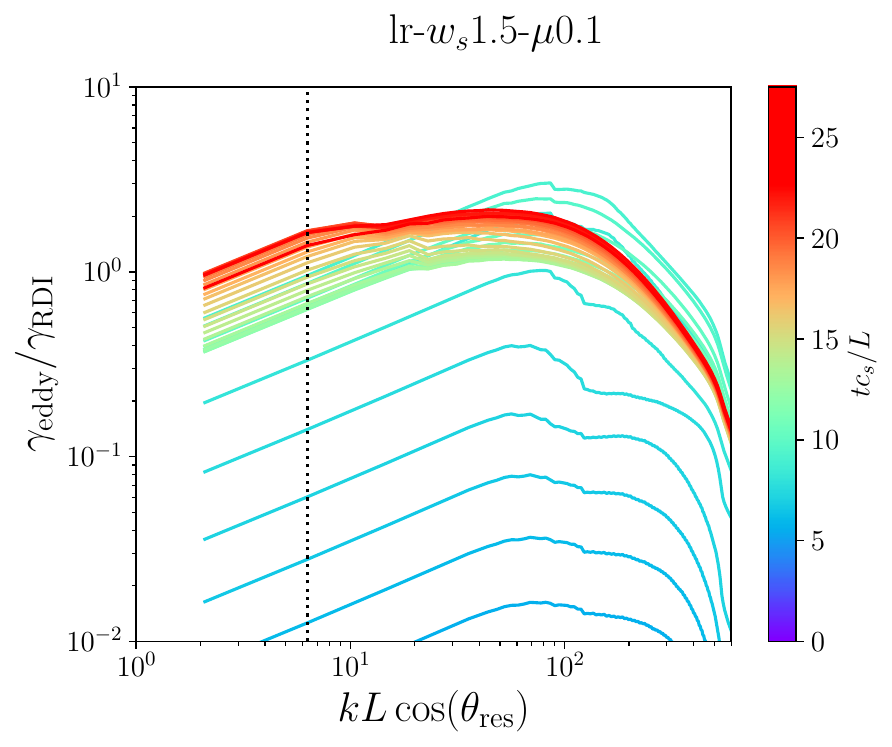}
    }
    \caption{The ratio of the eddy turnover rate (calculated per \cref{app:spectra}) to the linear instability growth rate (calculated analytically) as a function of wavenumber in a series of snapshots from the simulations discussed in \cref{sec:limits}.
    A vertical line is shown at (a) $kL\sin(\tres)=2\pi$ and (b) $kL\cos(\tres)=2\pi$, the minimum wavenumber that can support a resonant mode in this simulation geometry.}
    \label{fig:turnover_limits}
\end{figure}

While the majority of the simulations fit within the described picture of an acoustic RDI in balance with instability-driven turbulence across a forcing range, the behavior of two simulations, \siml{1.1} and \simu{1.5}, show marked deviations that warrant discussion.
As shown in \cref{fig:turnover_limits}, the ratio of eddy turnover rate to instability growth rate is order unity across a range comparable to the simulations in \cref{fig:turnover}, suggesting that the postulated balance holds. However, other diagnostics show behavior less consistent with the model described above.
We describe these discrepancies and propose possible causes.

\begin{figure*}
    \centering
    \includegraphics[width=0.8\textwidth]{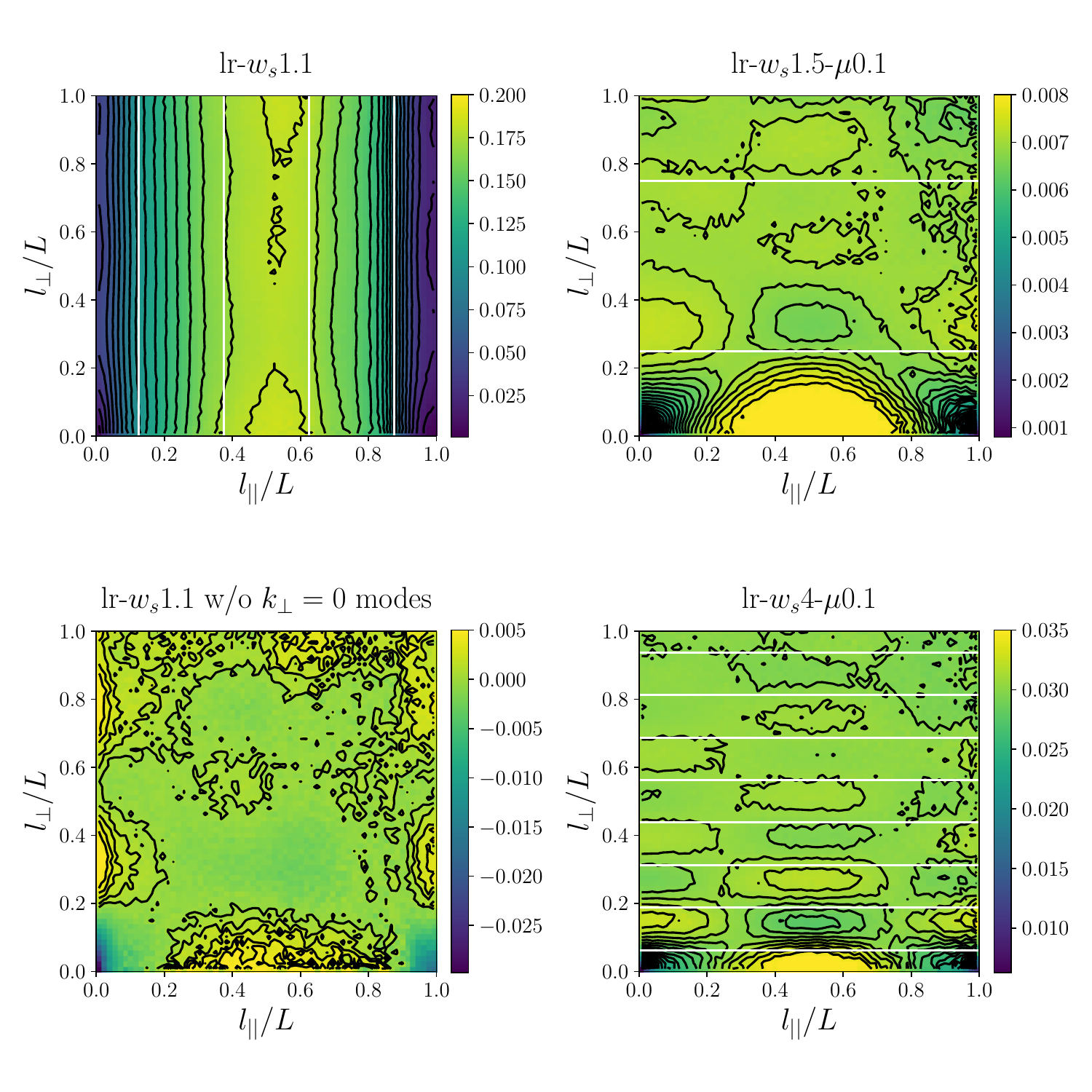}
    \caption{The two-dimensional second-order structure function of the gas density, as per \cref{fig:gas}, for the simulations discussed in \cref{sec:limits}.
    White lines indicate node lines for the expected outer scale perpendicular (or parallel in the case of \siml{1.1} since $\tres<\pi/4$) wave number as described in \cref{sec:largest}.
    At bottom left, the structure function for \siml{1.1} is plotted with the $\kperp=0$ modes removed.}
    \label{fig:gas-other}
\end{figure*}

\paragraph{\siml{1.1}}
The gas velocity and density statistics present in this simulation do not generally appear consistent with that of the other simulations.
While eddy turnover and instability growth rates appear to be commensurate across a range of scales (\cref{fig:turnover_limits}), the $k^{-2}$ gas velocity energy spectrum associated with this balance is not present (\cref{fig:1d-spectra} top row).
Further, in the saturated state, there is not clear evidence of large scale resonant modes.
This is visible in the gas density fluctuations (\cref{fig:rho_fourier} fifth row), which are not concentrated on the resonance cone as seen for other simulations.
It is also apparent in the gas density structure function.
Plotted as a two-dimensional function (as described in \cref{sec:largest}) in \cref{fig:gas-other}, the dominant mode has $\koperp =0$ and $\kopar=2\pi$.
This differs from the expected largest scale resonant mode, for which $\koperp L/2\pi=1$ and $\kopar L/2\pi=1/\tan(\tres)=2.18\approx 2$.
If this dominant mode is removed, by subtracting from the function its average over the $\kperp$ direction, effectively canceling any $\kperp=0$ modes, the result does not have a clear dominant mode (\cref{fig:gas-other} bottom left).
This is in marked contrast to the simulations in \cref{fig:gas}, which show clear large scale modes near the expected resonances.
The $\koperp =0$ and $\kopar=2\pi$ mode seen in \siml{1.1} essentially describes density stratification parallel to streaming in a periodic domain.
It may be associated with the shock-like structure visible in a snapshot of the gas velocity in the fifth row of \cref{fig:u_cross}, perhaps the result of the non-resonant \emph{pressure-free} modes discussed in \cref{sec:review} and in HS18.
Indeed, if the derivation in \cref{sec:spectra} is repeated assuming that the growth rate of the pressure-free mode ($\gamma_{PF}\sim k^{2/3}$) is in balance with eddy turnover, the spectrum would be expected to go as
\begin{equation}
E_\bu(k)\sim k^{-5/3},
\end{equation}
which is consistent with the slopes of the spectra with respect to $k$ and $k_\perp$ seen in \cref{fig:1d-spectra} for \siml{1.1}.
(It may also be argued that the $k^{-2}$ law expected for the acoustic RDI is in fact present in $E_\bu(\kpar)$ (\cref{fig:1d-spectra} top right) instead of $E_\bu(k)$ or $E_\bu(\kperp)$ due to the relatively shallow $\tres=\arccos(1/1.1)\approx 0.43$ for this low Mach number.)

\paragraph{\simu{1.5}}
Similarly, for this simulation, while the ratio of eddy turnover rate to instability growth rate is consistent with these timescales being in balance across an outer forcing range (\cref{fig:turnover_limits}), this does not result in the $k^{-2}$ gas velocity spectrum seen for comparable runs at lower $\mu$, \simh{1.5} and \siml{1.5} (\cref{fig:1d-spectra} second row).
However, unlike \siml{1.1}, the gas density structure function shows the clear presence of large scale modes (\cref{fig:gas-other}), albeit at a larger $\kperp$ than the expected resonant mode.

The deviation in behavior in these simulations may be attributable to a reduction in the efficacy of the acoustic RDI as a forcing mechanism at low streaming velocities, or to the presence of other forms of forcing.
For \siml{1.1}, in addition to the instability being weaker at lower streaming velocity, the velocity dispersion produced by the turbulence is sufficient that it may produce a trans-sonic regime that suppresses the resonant instability (which requires $w_s>c_s$ for resonance).
In particular (per \cref{tab:sims}) $u_{\rm rms}=0.33c_s$ at saturation, substantially larger than $w_s-c_s=0.1 c_s$. (Compare to \simh{1.5} with $u_{\rm rms}=0.41c_s$ at saturation and $w_s-c_s=0.5c_s$.)
In such circumstances, other non-resonant modes such as the pressure-free mode (HS18) that are linearly sub-dominant may be of greater importance, and we see evidence consistent with this mode for \siml{1.1}.

Turbulence may also be driven by the non-uniform drag force imparted on the gas by inhomogeneous dust distribution, which will be active regardless of the activity of the acoustic RDI in the saturated state.
The drag force on the gas is $\propto\mu$, such that the driving of the turbulence by dust inhomogeneity might naively be expected to scale linearly with $\mu$, which is stronger than the scaling of the instability $\grdi\sim\mu^{1/2}$.
In addition, per \cref{tab:sims}, the dust density fluctuation amplitudes (and presumably therefore the strength of stirring) for the $\mu=0.1$ simulations are dramatically larger than for comparable $\mu=0.01$ simulations, in both absolute terms and when scaled to the average dust density.
These simulations also show channel structures in the gas velocity (\cref{fig:u_cross}), which may help explain the anomalous behavior of \simu{1.5}.
As in \cite{moseleyNonlinearEvolutionInstabilities2019}, these may be caused by a combination of dust clumping and the periodic boundary conditions.
As the forcing by dust drag occurs at a range of scales dependent on the clumping behavior of the dust, it would produce a different spectrum than forcing by the acoustic RDI, which may help explain the differences seen between \simu{1.5} and \siml{1.5} or \simh{1.5} in the second row of \cref{fig:1d-spectra}.
It is worth noting that the presence of balanced timescales (\cref{fig:gas-other}) without the associated $k^{-2}$ spectra is not wholly contradictory, as the scale ranges in which the ratio of timescales is order unity for other simulations (\cref{fig:gas}) extend further than the ranges in which those simulations show $k^{-2}$ spectra (\cref{fig:1d-spectra}).

\begin{figure}
    \centering
    \includegraphics[width=\linewidth]{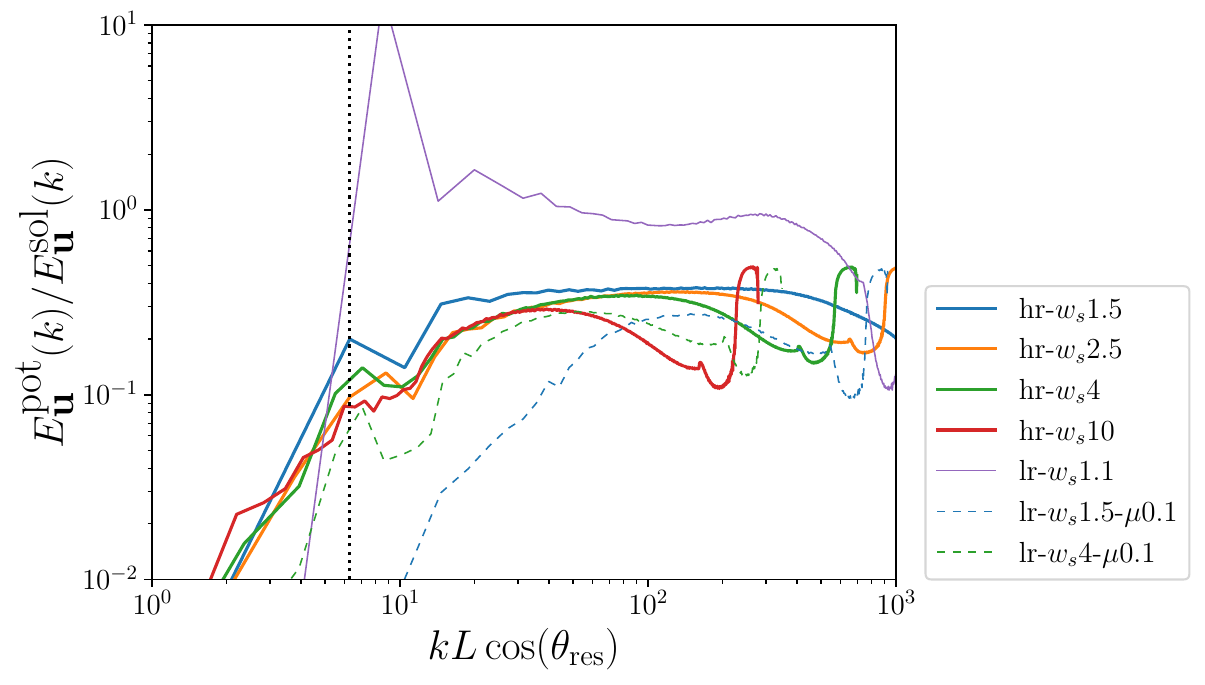}
    \caption{Ratio of kinetic energy spectra of the potential and solenoidal components of the gas velocity for various simulations.
    A vertical line has been placed at $kL\cos(\tres)=2\pi$ to indicate the largest scale at which resonant modes are supported by the simulation geometry for $w_s/c_s\geq\sqrt{2}$.}
    \label{fig:potsol}
\end{figure}

A qualitative difference in dynamics between these simulations and the others is further emphasized by the distribution of kinetic energy between the potential and solenoidal components of the gas flow.
This can be seen in the ratios of the potential and solenoidal spectra shown in \cref{fig:potsol}.
While this ratio is largely in the range $0.1-0.3$ for scales between the outer and dissipative scales for most simulations, it is dramatically larger ($\gtrsim1$) for \siml{1.1}, and both smaller and highly variable for \simu{1.5}.
In both cases, this may be related to the mechanisms described above.
In particular, the concentration of energy in potential flow in \siml{1.1}, with a large spike near the box scale, may be related to the shock-like longitudinal discontinuity in velocity seen in \cref{fig:u_cross}.
In addition, the weighting of kinetic energy towards solenoidal flow in \simu{1.5} would be consistent with the inhomogeneous mixing by dust mentioned above.
While this is by no means conclusive, it emphasizes the presence of complex and parameter-dependent dynamics in these systems, which may occur alongside and interact with the resonant instabilities we set out to study.

Interestingly, another case with $\mu=0.1$ but with higher streaming velocity, \simu{4}, can be seen to have behavior more closely resembling the $\mu=0.01$ case. (Compare to \simh{4} and \siml{4}.)
Forcing at an outer scale by balanced resonant unstable modes is visible in gas density fluctuations (\cref{fig:rho_fourier} bottom row), spectra (\cref{fig:1d-spectra} fifth row), and $S_2^\rho$ (\cref{fig:gas-other}).
This suggests a complex dependence on streaming velocity and dust loading, the mapping of which is left for future work.

%% file: sections/8conclusion.tex
\section{Discussion}
\label{sec:conclusion}

\subsection{Summary of results}

These results consist of the first investigation focused on the detailed saturation mechanisms of the mid-$k$ acoustic RDI, and the highest resolution simulations of the acoustic RDI (isolated from other instabilities) published to date.
Our primary results consist of a model of the saturation process, which is able to predict gas velocity fluctuation amplitudes and energy spectra of the saturated state, and numerical evidence of its validity across a range of dust streaming velocities.
The initial linear growth phase is seen to be dominated by the smallest well resolved scales, as these have the fastest growth.
As these saturate due to self-excited turbulent fluctuations, larger scales catch up and saturate in similar fashion.
This process halts when it reaches the slowest-growing modes near the box scale, producing a stationary turbulent state.
These results extend those of \citet{moseleyNonlinearEvolutionInstabilities2019}, and by producing analogous phenomena using a substantially different code, demonstrate the robustness of our modeling of the RDI as physically meaningful and independent of numerical scheme.

Following saturation, the instability remains active near the box scale, limited by the turbulence it produces.
This is evident in a $k^{-2}$ velocity power spectrum at this scale, indicative of a balance between the eddy turnover rate and the instability growth rate.
Within the forcing range, the turbulence is strongly anisotropic due to the anisotropy of the resonant unstable modes.
This  range likely does not extend indefinitely toward smaller scale, as the growth rate of the acoustic RDI is expected to scale more weakly with $k$ than the eddy turnover rate of the turbulent cascade.
As such, at sufficiently small scales, the dynamics would be dominated by a turbulent cascade of energy from larger scales, with the instability suppressed.
This produces an inertial range, the outer edge of which is resolvable at the available simulation resolution.
As the only characteristic scales are the outer and dissipation scales, universal behavior independent of the character of the forcing might be expected (within a sufficiently wide inertial range).
We find evidence consistent with the presence of this regime.

The characteristics of the turbulence are seen to be dependent on the streaming velocity and on the dust to gas mass density ratio.
In particular, the width of the forcing range decreases as streaming velocity increases.
This may be attributed to the stronger forcing produced by higher streaming velocities, which results in larger eddy velocities and more effective suppression of the instability.
In addition, at sufficiently low streaming velocities and sufficiently large dust loading, various features of the saturated state change markedly.
We suggest that this may be due to the contribution of non-resonant forcing mechanisms.

These insights into the saturated acoustic RDI are of both theoretical interest and direct relevance to astrophysical phenomena.
By describing and demonstrating the mechanisms by which acoustic RDIs saturate, it offers a framework to guide further study of this and other resonant drag instabilities.
In particular, the resemblance of the acoustic RDI to the slow and fast magnetosonic RDIs suggests the possibility that similar mechanisms might be at play in the saturation of those instabilities \citep{hopkinsUbiquitousInstabilitiesDust2018,hopkinsResonantDragInstability2018}.
In addition, as the acoustic RDI has already been shown to be unstable in conditions relevant to astrophysical environments such as active galactic nuclei and cool stellar winds, the physics described above may play a role in shaping turbulent flows in these environments \citep{hopkinsUbiquitousInstabilitiesDust2018,hopkinsResonantDragInstability2018,israeliResonantInstabilitiesMediated2023a}.

\subsection{Further work}

Several directions for further investigation immediately suggest themselves.
First, the above analysis does not consider the statistics of the dust in detail.
This existing simulation dataset lends itself to a study of the behavior of the dust, including evolution through the growth and saturation phases, clumping and structure formation, and density and velocity correlation with the gas.
This could yield further interesting insights into the nonlinear saturation of this instability and its role in dust evolution within astrophysical environments.
In particular, it may shed light on the role of resonant and non-resonant forcing mechanisms speculated upon in \cref{sec:limits}.

In addition, there are clear areas in which further simulations might refine these existing results.
The ability of these simulations to resolve the behavior of the inertial range, particularly slopes of spectra, is limited simply by resolution.
Higher resolution simulations, or ones using codes capable of handling non-cubic domains, would allow a more robust range of scales to be simultaneously modelled.
Simulations may also be performed that incorporate the high-$k$ or low-$k$ regimes, and perhaps their interaction with the mid-$k$ regime studied here.
A larger set of simulations, comprehensively scanning parameter space, would be necessary to fully map the changes in behavior with varied streaming velocity and dust loading touched upon here.

Further, predictive analytic models of the saturated statistics of acoustic RDIs remain incomplete.
An explanation of the parameter dependence of the saturated state, including the width of the forcing range and slope of the parallel spectra, is not yet available.
Theoretical study of the dust statistics, particularly clumping behavior and velocity dispersion, would be particularly valuable for models of dust formation, destruction, and agglomeration in astrophysical systems~\citep{pumirCollisionalAggregationDue2016}.

Separately, the study of a number of other RDIs is also at an early stage.
A study resembling this one, with the inclusion of dust charge and magnetic fields, could yield valuable insight into the saturation and balance mechanisms of MHD RDIs, and the character of the resulting turbulence.
While the nonlinear saturation of MHD RDIs has been studied \citep{seligmanNonlinearEvolutionResonant2019,hopkinsSimulatingDiverseInstabilities2020}, theoretical models remain incomplete.
As such, by presenting insights into fundamental mechanisms underlying the nonlinear evolution of RDIs, this work may help build a foundation for the ongoing investigation of this complex set of related phenomena.
As these instabilities are of interest in the modelling of dust in diverse processes in astrophysics, ranging in scale from the dynamics of active galactic nuclei to the formation of planets, developing a thorough understanding of their nonlinear behavior remains a challenge of substantial importance.

%% file: sections/9acknowledgments.tex
\section*{Acknowledgments}
We would like to thank Philip F. Hopkins for useful conversations concerning this work.
This work was supported by the U.S. Department of Energy, Office of Science, Office of Fusion Energy Sciences, and has been authored by Princeton University under Contract Number DE-AC02-09CH11466 with the U.S. Department of Energy.
JS acknowledges the support of the Royal Society Te Ap\=arangi, through Marsden-Fund grant MFP-UOO2221  and  Rutherford Discovery Fellowship RDF-U001804.

\section*{Data Availability}

The simulation data presented in this article is available upon request to BYI.

%% file: sections/Abenchmark.tex
\section{Benchmarking RAMSES for simulation of acoustic RDIs}
\label{app:benchmark}

In order to confirm that RAMSES accurately models acoustic RDIs, a set of benchmark tests of the linear growth of individual RDI modes was performed.
The simulations were initialized with conditions described at the beginning of Section~\ref{sec:sim-params}, with the modification that the gas and dust densities and velocities were seeded with a small inhomogeneity, corresponding to a resonant eigenmode of the linearized system~\citep{hopkinsResonantDragInstability2018,squireResonantDragInstability2018}.
In particular, we use the fastest growing mode at the resonant angle, the resonant quasi-drift mode.
The magnitude of this inhomogeneity was set by multiplying a unit eigenvector (with densities normalized with respect to their equilibrium values and velocities normalized with respect to the sound speed $c_s$) by a factor of $10^{-3}$.
This inhomogeneity can be written as
\begin{equation}
\begin{split}
(\frac{\delta \rho_d}{\rho_{d0}},\ \frac{\delta \bv}{c_s},\ \frac{\delta \rho}{\rho_{0}},\ \frac{\delta \bu}{c_s})=&10^{-3}\times\hat V \exp[i(k_x x+k_z z)]\\
=&10^{-3}\times\hat V \exp\left\{ik_z\left[\left(\frac{w_s^2}{c_s^2}-1\right)^{1/2} x+z\right]\right\}
\end{split}
\end{equation}
where $\hat V$ is the unit eigenvector of the linearized system corresponding to the quasi-drift mode.

The streaming velocity was chosen to be $w_s=\sqrt{2}c_s\approx1.41 c_s$ so that the resonant angle $\theta=\pi/4$ allowed precisely resonant modes to be compatible with the grid and periodic boundary conditions.
The mass density ratio and dust stopping time were set at $\mu=0.1$ and $t_s=0.1L/c_s$.
Various values of the wave number for the seeded inhomogeneity were used in order to gauge simulation accuracy across scales: $k_zL/2\pi=2,\ 4,\ 8,\ 16$.
Tests were performed on $64^3$, $128^3$, and $256^3$ grids to determine scaling with respect to resolution.

\begin{figure}
\centering
\subfloat[]{
\includegraphics[width=0.95\linewidth]{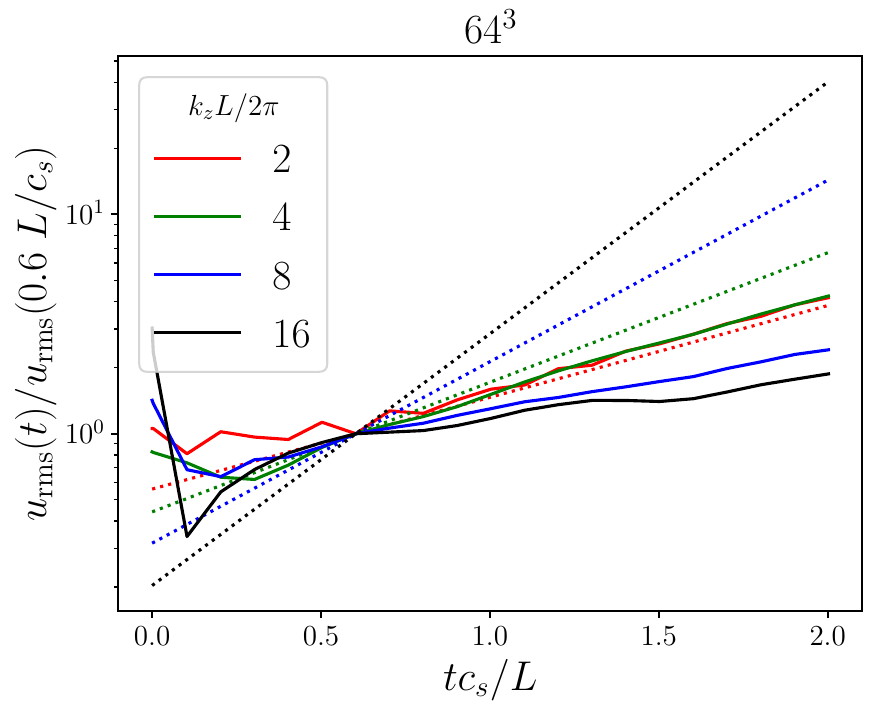}
}

\subfloat[]{
\includegraphics[width=0.95\linewidth]{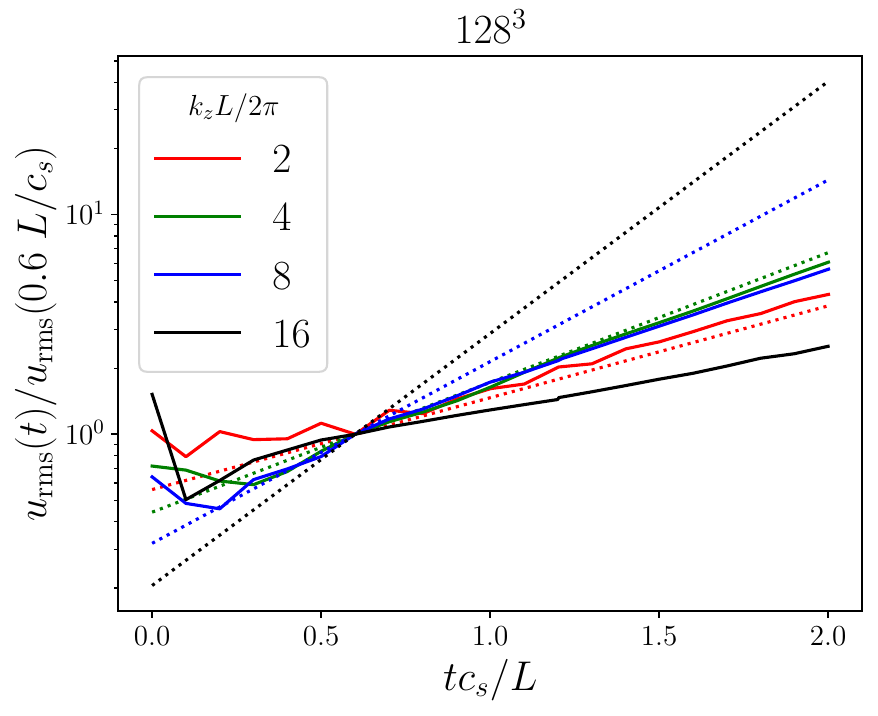}
}

\subfloat[]{
\includegraphics[width=0.95\linewidth]{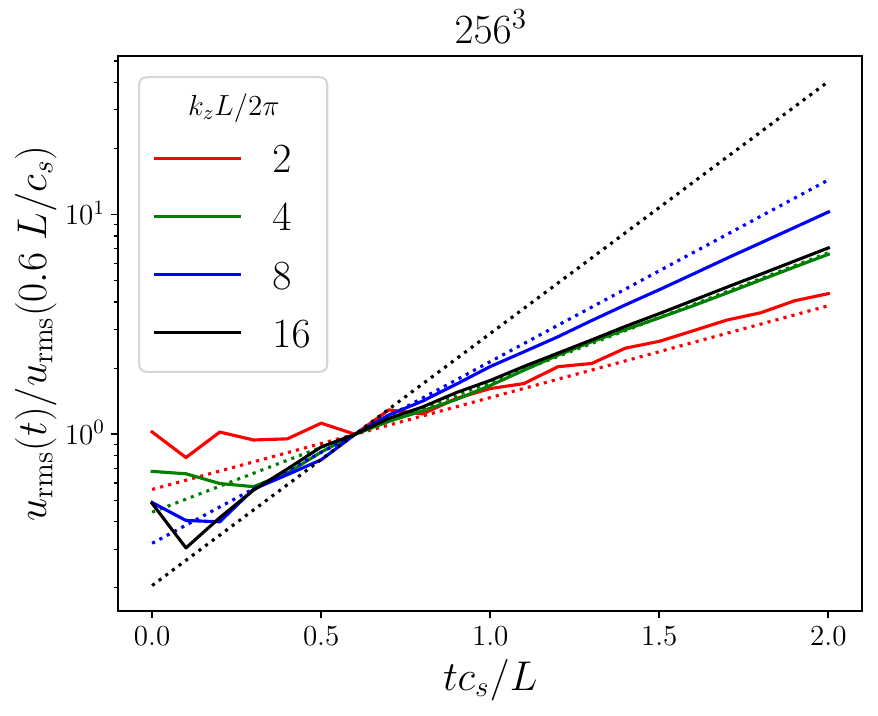}
}

\caption{Normalized root mean squared gas velocity fluctuations over time for the linear growth of a seeded acoustic RDI instability for varied wavelength at resolutions of (a) $64^3$, (b) $128^3$, and (c) $256^3$. Dashed lines indicate the analytically calculated growth rates.}
\label{fig:bench}
\end{figure}

\Cref{fig:bench} shows the evolution of the simulations, and \cref{tab:bench} lists the corresponding growth rates.
At large scales, the simulations reproduce the expected linear growth rate within reasonable accuracy.
As shown in \cref{tab:bench_norm}, for a simulation with $n^3$ gridpoints, accuracy begins to break down at $k_zL/2\pi=n/16$, and the instability is substantially damped for $k_zL/2\pi\geq n/8$.
In other words, the simulations accurately describe unstable modes with wavelengths longer than $\sim16$ gridpoints.
Below this scale, simulated growth rates are slower than analytic predictions, perhaps due to numerical viscosity.
A similar test was performed in \cite{squireAcousticResonantDrag2022} using the GIZMO code, also finding quantitative agreement that deteriorated at moderately small scales.

\begin{table}
\centering
\begin{tabular}{|c|c|c|c|c|}
\hline
$k_zL/2\pi$ & \text{analytic} & $64^3$ & $128^3$ & $256^3$ \\
\hline
2  & 0.96 & 0.98 & 1.01 & 1.02 \\
4  & 1.36 & 1.03 & 1.28 & 1.35 \\
8  & 1.90 & 0.63 & 1.20 & 1.63 \\
16 & 2.64 & 0.44 & 0.66 & 1.38 \\
\hline
\end{tabular}
\caption{Growth rates in units of $L/c_s$ for the acoustic RDI at each wavelength tested in the seeded benchmark tests, calculated analytically and from simulations at the specified resolutions. The growth rate was calculated from simulation data by least squares for points between $tc_s/L=0.6$ (the intercept point in Figure~\ref{fig:bench}) and $tc_s/L=2$ (the end of the simulation).}
\label{tab:bench}
\end{table}

\begin{table}
\centering
\begin{tabular}{|c|c|c|c|}
\hline
$2\pi n/k_zL$ & $64^3$ & $128^3$ & $256^3$ \\
\hline
4 & 0.17 &   &   \\
8 & 0.33 & 0.25 &   \\
16 & 0.75 & 0.63 & 0.52 \\
32 & 1.02 & 0.94 & 0.86 \\
64 &   & 1.05 & 0.99 \\
128 &   &   & 1.05 \\
\hline
\end{tabular}
\caption{Growth rates for the instability from simulations per \cref{tab:bench}, normalized with respect to their analytically calculated values.
They have been listed in order of number of gridpoints per wavelength, in order to demonstrate that the simulations faithfully replicated instabilities with wavelengths longer than $\sim 16$ gridpoints.}
\label{tab:bench_norm}
\end{table}

%% file: sections/Bspectra.tex
\section{Defining energy spectra and eddy turnover rate}
\label{app:spectra}
Energy spectra of the gas velocity are a key diagnostic of the turbulence studied in this work, and are used as a means of estimating the eddy turnover rate as a function of scale. The spectrum with respect to $k$ (direction-independent) is defined as
\begin{equation}
E_\bu(k)=\frac12\int d\phi d\theta k^2 \sin(\phi) \left|\tilde \bu(\bk)\right|^2,
\label{equ:spec-iso}
\end{equation}
where $\tilde \bu(\bk)$ denotes the Fourier transform of $\bu(\br)$, and we have used spherical coordinates $(k,\phi,\theta)$ for $\bk$.
We will also use the perpendicular and parallel energy spectra,
\begin{equation}
\begin{split}
E_\bu(k_\perp)=&\frac12\int dk_{||} d\theta k_\perp \left|\tilde \bu(\bk)\right|^2,\\
E_\bu(k_{||})=&\frac12\int dk_\perp d\theta k_\perp \left|\tilde \bu(\bk)\right|^2,
\end{split}
\label{equ:spec-aniso}
\end{equation}
where we have used cylindrical coordinates $(k_\perp,\theta,k_{||})$ aligned with $\hat{z}$, the streaming direction for $\bk$.
\label{app:turnover}

The eddy turnover rate for eddies of size $k$ may be estimated as
\begin{equation}
\geddy(k)=\frac k\pi\left[S_2\left(\frac{\pi}{k}\right)\right]^{1/2},
\end{equation}
where
\begin{equation}
\label{equ:s2}
S_2(l)=\left<\left[\bu(\br+l \hat{\bl})-\bu(\br)\right]^2\right>
\end{equation}
(averaged over all $\br$ and spherically over all unit vectors $\hat{\bl}$) is the second-order structure function of the gas velocity.
This may be associated with eddy turnover, as it describes the velocity increment across a particular scale, i.e. the typical velocity of an eddy of that scale.
This in turn may be approximated \citep{davidsonOriginsNatureTurbulence2015} as
\begin{equation}
S_2\left(\frac{\pi}{k}\right)\approx\frac43 \int_k^\infty dk E_\bu(k).
\label{equ:turb-s2}
\end{equation}
Putting these together, we obtain
\begin{equation}
\label{equ:eddy_expression}
\geddy(k)\approx \frac k\pi \left[\frac43 \int dk E_\bu(k)\right]^{1/2} \sim k^{3/2} \left[E_\bu(k)\right]^{1/2}.
\end{equation}